\documentclass{article} 
\usepackage{iclr2026_conference}

\usepackage{amsmath,amsfonts,bm}

\def\eqref#1{equation~\ref{#1}}

\def\1{\bm{1}}

\DeclareMathAlphabet{\mathsfit}{\encodingdefault}{\sfdefault}{m}{sl}
\SetMathAlphabet{\mathsfit}{bold}{\encodingdefault}{\sfdefault}{bx}{n}

\usepackage{microtype}
\usepackage{hyperref}
\usepackage{url}
\usepackage{graphicx}
\usepackage{booktabs}
\usepackage{subcaption}
\usepackage{xcolor}
\usepackage{multirow}
\usepackage{amsmath,amssymb}
\usepackage{wrapfig}
\usepackage{colortbl}
\usepackage{tikz}
\usetikzlibrary{arrows.meta, positioning, shapes.geometric, fit, calc, backgrounds}

\usepackage{times}
\usepackage{latexsym}
\usepackage{booktabs}
\usepackage{float}
\usepackage{pifont}
\usepackage{xcolor}
\usepackage{xcolor}
\usepackage{xspace}
\usepackage{graphicx}
\usepackage{multirow}
\usepackage{tabularx}
\usepackage{tabularx}
\usepackage{makecell}
\usepackage{algorithm}
\usepackage{algpseudocode}
\usepackage{wrapfig}
\usepackage{longtable}
\usepackage[most]{tcolorbox} 
\usepackage{xcolor}      
\usepackage{enumitem}
\usepackage{CJKutf8}
\usepackage{amssymb}
\usepackage{caption}

\usepackage{xcolor,pifont}
\definecolor{mygreen}{RGB}{34,139,34}

\usepackage{alltt}
\usepackage{xspace}
\usepackage{gradient-text}
\usepackage{lineno}

\newcommand{\bench}{{\fontfamily{pbk}\selectfont \gradientRGB{GBQA}{25, 24, 59}{25, 24, 59}}\xspace}
\newcommand{\castle}{{\fontfamily{pbk}\selectfont \gradientRGB{CASTLE}{25, 24, 59}{25, 24, 59}}\xspace}

\definecolor{lightblue}{RGB}{220,235,252}
\definecolor{lightgray}{RGB}{242,242,242}

\definecolor{bugeasy}{RGB}{0, 128, 0}
\definecolor{bugmedium}{RGB}{204, 102, 0}
\definecolor{bughard}{RGB}{180, 0, 0}
\newcommand{\diffeasy}[1]{\textcolor{bugeasy}{\textbf{#1}}}
\newcommand{\diffmedium}[1]{\textcolor{bugmedium}{\textbf{#1}}}
\newcommand{\diffhard}[1]{\textcolor{bughard}{\textbf{#1}}}

\title{
\raisebox{-0.5ex}{\includegraphics[height=2.2em]{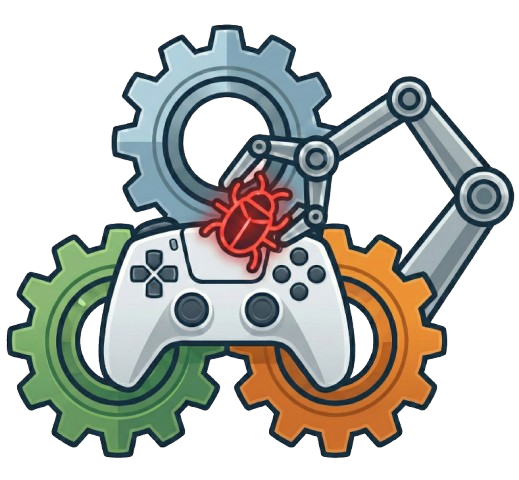}}
\bench: A Game Benchmark for Evaluating LLMs as Quality Assurance Engineers}

\usepackage{authblk}
\newcommand{\aspace}{\hspace{1em}}

\newcommand{\HKU}{$^1$}
\newcommand{\Independent}{$^2$}
\newcommand{\WESTLAKE}{$^3$}
\newcommand{\Datawhale}{$^4$}

\author{
\textbf{Shufan Jiang}\HKU\textsuperscript{,}\Datawhale\aspace
\textbf{Chios Chen}\Independent\aspace
\textbf{Zhiyang Chen}\thanks{Corresponding to \href{mailto:volgachen@gmail.com}{volgachen@gmail.com}}\hspace{4.5pt}\WESTLAKE

\vspace{0.3em}

\begin{footnotesize}
\HKU The University of Hong Kong \aspace
\Independent Independent Researcher \aspace
\WESTLAKE Westlake University \aspace
\Datawhale Datawhale Org.
\end{footnotesize}
}

\iclrfinalcopy 

\begin{document}


\maketitle

\begin{abstract}
    The autonomous discovery of bugs remains a significant challenge in modern software development. Compared to code generation, the complexity of dynamic runtime environments makes bug discovery considerably harder for large language models (LLMs).
    In this paper, we take game development as a representative domain and introduce the Game Benchmark for Quality Assurance (\bench), a benchmark containing 30 games and 124 human-verified bugs across three difficulty levels, to evaluate whether LLMs can autonomously detect software bugs.
    The benchmark is constructed using a multi-agent system that develops games and injects bugs in a scalable manner, with human experts in the loop to ensure correctness.
    Moreover, we provide a baseline interactive agent equipped with a multi-round ReAct loop and a memory mechanism, enabling long-horizon exploration of game environments for bug detection across different LLMs.
    Extensive experiments on frontier LLMs demonstrate that autonomous bug discovery remains highly challenging: the best-performing model, Claude-4.6-Opus in thinking mode, identifies only 48.39\% of the verified bugs.
    We believe \bench provides an adequate testbed and evaluation criterion, and that further progress on it will help close the gap in autonomous software engineering.
\end{abstract}

\section{Introduction}

\begin{figure}[t]
\centering
\includegraphics[width=1\textwidth]{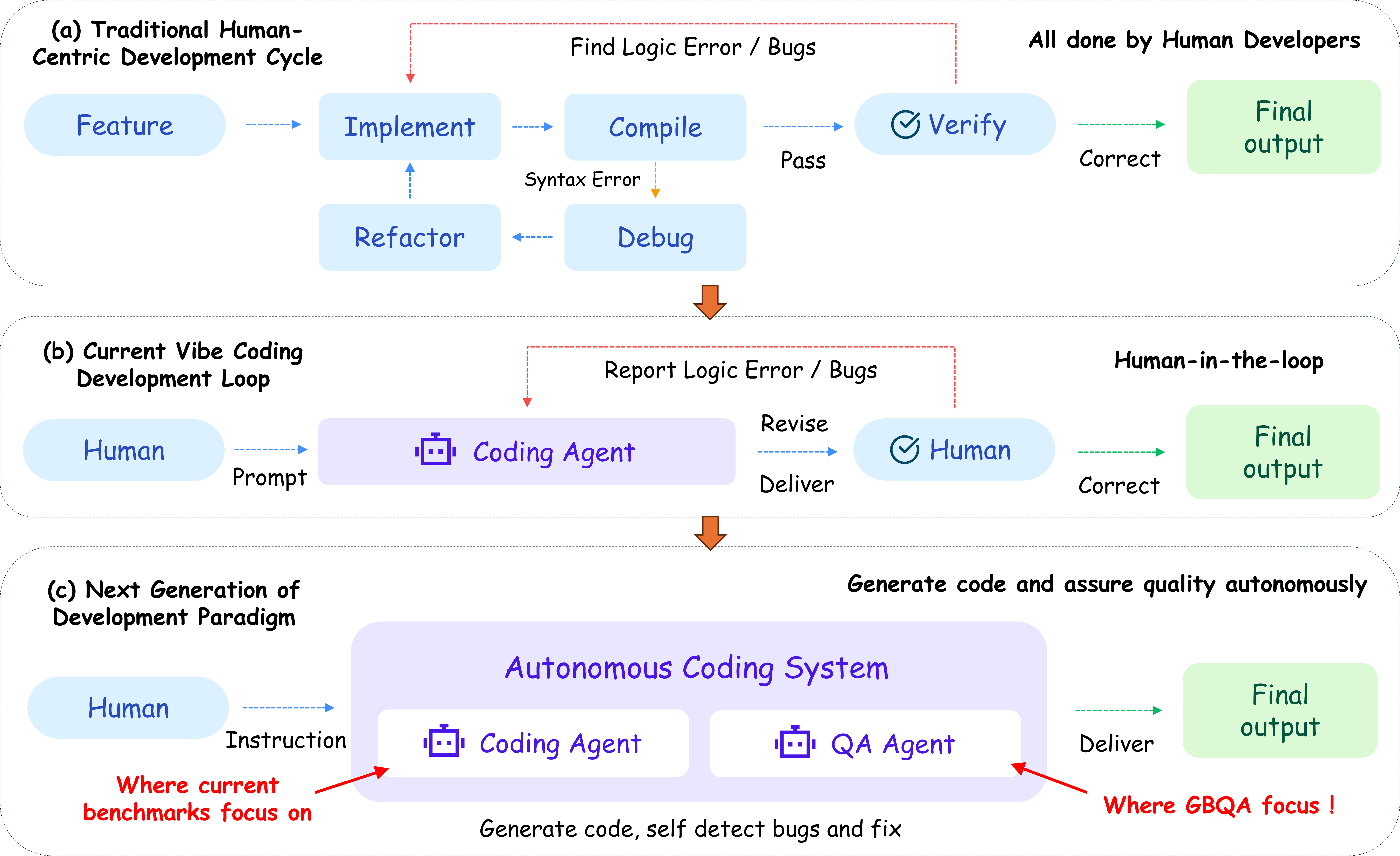}
\caption{Evolution of the software development paradigm in the LLM era.
(a) Traditional human-driven iterative workflow.
(b) Human--LLM collaborative coding, where a coding agent assists development under human supervision.
(c) Toward a fully autonomous coding system which can generate code, detect bugs and fix them without human-in-the-loop.
While existing benchmarks primarily focus on code generation and fixing, our benchmark emphasizes autonomous bug discovery and quality assurance part within the development cycle.}
\label{fig:dev}
\end{figure}

Real-world software development is systematic: no non-trivial system is correct and robust on the first attempt, thus requiring an inherently iterative software engineering workflow.
Traditionally, human developers follow repeated cycles of implementation, testing, debugging, and refactoring as shown in Figure~\ref{fig:dev}(a).
Currently, coding agents like Claude Code~\citep{anthropic2025claudecode}, Cursor~\citep{cursor2024editor}, and OpenAI Codex~\citep{openai2025codex}, actively participate in this development loop.
In this paradigm, human developers provide natural language instructions, while LLMs generate code, execute the resulting program, inspect failures, and iteratively revise the code.
This workflow, often referred to as vibe coding~\citep{karpathy2025vibecoding}, pushes the frontier of automatic software engineering as illustrated in Figure~\ref{fig:dev}(b) and (c).

Within this classical cycle, recent progress has dramatically strengthened the \textbf{development and fixing} side.
Frontier LLMs can now generate project-level codebases from natural language specifications~\citep{qian2024chatdev, hong2023metagpt} and resolve real-world code issues given well-written bug reports or issue descriptions~\citep{jimenez2024swebench, xia2024agentless}.
However, the \textbf{testing and bug discovery} side of this loop remains largely unexplored, upon which the quality of a released software critically depends.

Unlike code generation or fixing, bug discovery poses fundamentally different challenges.
First, the objective is ill-defined: the agent must proactively determine that ``something is wrong'' without being told what to look for, unlike generation or fixing tasks where a clear target or issue description is provided.
Second, effective bug discovery demands comprehensive exploration and systematic planning over large behavioral state spaces, rather than targeted edits to a known location.
Third, the agent must reason about the gap between expected and actual runtime behavior, often without access to explicit specifications.
Most existing benchmarks bypass these difficulties by articulating a precise description in the task description before the agent intervenes. Consequently, the cognitively demanding work of perceiving anomalies and localizing their causes is still completed by humans.
This upstream gap is similarly highlighted by recent efforts in autonomous code auditing~\citep{guo2025repoaudit} and large-scale bug mining~\citep{bugstone2025}.
Advancing toward fully autonomous system, therefore, requires directly evaluating and improving the ability of LLMs to discover defects independently.

In this paper, we take game development as the testbed for autonomous bug discovery.
Games are self-contained software systems composed of internal state management, user input handling, and output rendering.
They require long-term dynamic interactions within a single session, making them ideal representatives of real-world software engineering settings.
At the same time, games expose clearly defined action spaces and state transitions, making agents easily construct formatted inputs and outputs, naturally compatible with agent-based exploration.
Such interaction-driven, stateful verification is precisely the agentic capability that next-generation LLMs need to develop.
Moreover, bug discovery in games corresponds to Quality Assurance~(QA) in real world applications, which has a long tradition of systematic and specification-driven testing~\citep{Myers1979ArtOS, ammann2016introduction}.

\begin{figure}[t]
\centering
\includegraphics[width=1\textwidth]{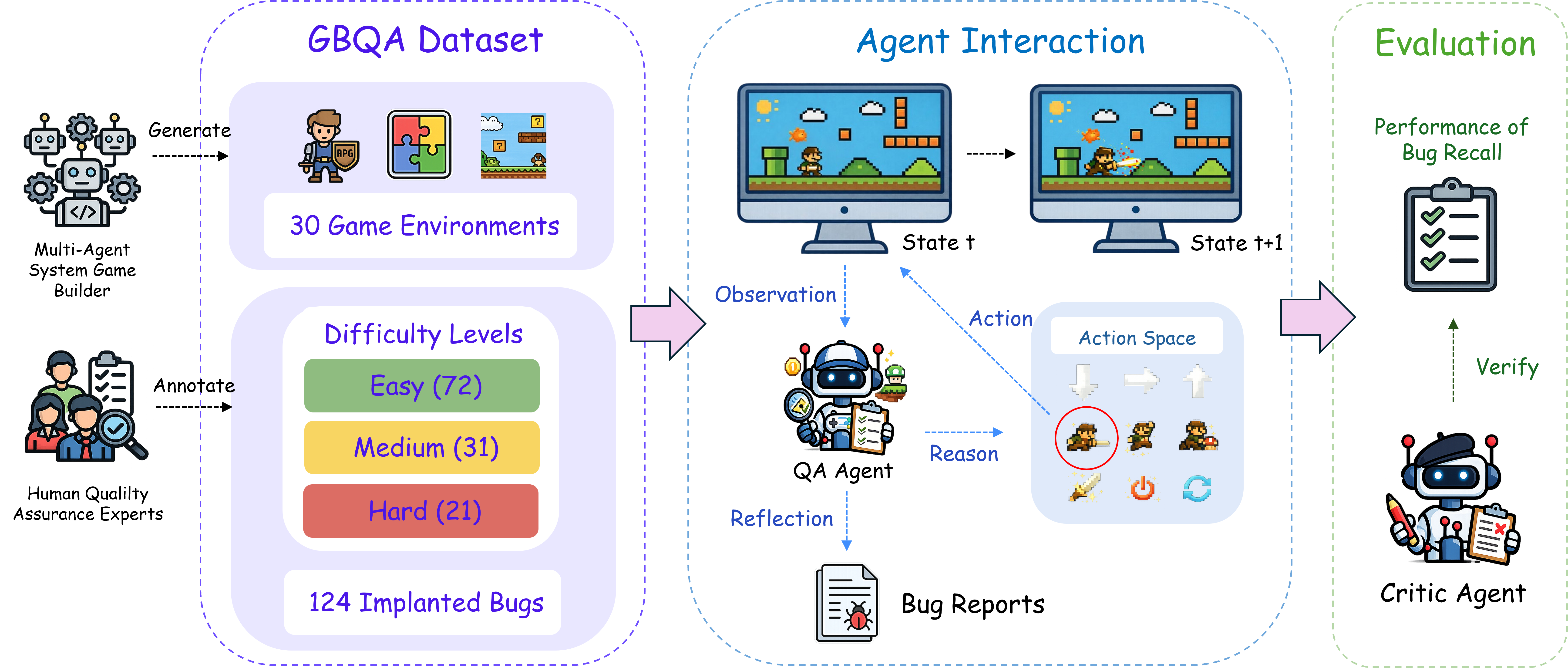}
\caption{
Overview of \bench. Dataset is constructed using a multi-agent game builder that generates 30 game environments with 124 implanted bugs, which are annotated and categorized into three difficulty levels (Easy, Medium, Hard) by human QA experts. During evaluation, a QA agent autonomously interacts with the game environment through ReAct loops, and produces structured bug reports. Then, a critic agent verifies reported bugs by matching them against human-annotated ground truth to compute quantitative metrics. 
}
\label{fig:benchmark}
\end{figure}

Motivated by these considerations, we introduce \bench, a benchmark designed to evaluate the ability of LLMs to autonomously discover bugs in interactive game environments.
As illustrated in Figure~\ref{fig:benchmark}, \bench contains 30 diverse games with a total of 124 human-verified bugs in them in order to evaluate how well an agent performs in the QA task. During evaluation, the agent autonomously explore the games, identify potential bugs, and report clear descriptions along with reproducible steps. Subsequently, each reported bug is then matched against the human-verified ground-truth annotations to compute quantitative metrics. The annotated bugs are categorized into different difficulty levels to assess model robustness across varying complexity.
To construct this benchmark at scale, we develop a multi-agent system that automatically generates games and injects bugs with controllable complexity, while human experts remain in the loop to verify the correctness of all annotations.

To evaluate the capability of frontier LLMs in bug detection, we further provide a baseline interactive agent equipped with a multi-round ReAct loop and a memory mechanism, enabling long-horizon exploration of game environments.
Our experiments demonstrate that autonomous bug discovery remains highly challenging: even the best-performing model, Claude-4.6-Opus in thinking mode, identifies less than half of the bugs, revealing substantial room for improvement.

Our contributions can be summarized as follows.
\begin{itemize}[leftmargin=*,labelsep=2mm]
    \item We formalize the problem of autonomous bug discovery in interactive environments and present \bench, a benchmark containing 30 diverse games and 124 human-verified bugs across three difficulty levels, along with a critic agent that supports automated evaluation.
    \item We develop a scalable game environment builder, including a multi-agent system capable of generating games and inserting bugs with controllable complexity, and introduce human-in-the-loop to ensure its correctness. 
    \item We perform extensive evaluations of cutting-edge LLMs in \bench, providing not only a comprehensive analysis of their performance and limitations but also a characterization of current failure modes in autonomous bug discovery.
\end{itemize}

\section{Related Work}

\textbf{Software Engineering and Agent Benchmarks.}
Large language models have been widely evaluated on software engineering tasks.
SWE-bench~\citep{jimenez2024swebench} and its extensions~\citep{aleithan2024swebenchenhancedcodingbenchmark} measure an agent’s ability to resolve real-world GitHub issues, while systems such as Agentless~\citep{xia2024agentless} improve issue localization and patch generation via structured pipelines.
A shared assumption across these benchmarks is that the bug has already been identified and described by humans; agents are evaluated primarily on code repair.
Beyond issue-driven repair, recent work explores automated defect detection in static repositories.
RepoAudit~\citep{guo2025repoaudit} and BugStone~\citep{bugstone2025} analyze structural patterns and data dependencies to discover vulnerabilities at scale.
However, these approaches operate on static code and do not assess the ability of agent to interact with a dynamic system, execute multi-step behaviors, and infer specification-level inconsistencies from runtime feedback.
More generally, interactive agent benchmarks such as WebArena~\citep{zhou2024webarena} and AgentBench~\citep{liu2024agentbench} evaluate LLM agents in web navigation and tool-use scenarios.
SMART~\citep{mu2025synergizingcodecoveragegameplay} incorporates coverage-aware strategies for functional testing.
In these settings, the environment is treated as ground truth and success is defined by task completion.
In contrast, \bench treats the environment itself as the object of evaluation and introduces flaw discovery rate as a complementary metric for agentic software engineering.

\textbf{Game-Based Agents and Automated Game Testing.}
Interactive games have become a major testbed for LLM agents.
Voyager~\citep{wang2023voyager}, MineDojo~\citep{fan2022minedojo}, CRADLE~\citep{tan2024cradle}, and Generative Agents~\citep{park2023generative} focus on goal achievement and skill acquisition in correctly functioning environments.
Closer to our setting, TITAN~\citep{titan2025} and Orak~\citep{orak2025} explore LLM assisted game testing.
While demonstrating the feasibility of QA-oriented agents, these systems operate in proprietary environments without publicly verifiable bug annotations, limiting standardized comparison.
\bench differs in two respects:
(1) it provides fully known and human-verified bug annotations, enabling rigorous quantitative evaluation; and
(2) it offers a scalable environment builder that supports controllable complexity and systematic benchmark expansion.
Together, these properties establish \bench as a standardized testbed for autonomous bug discovery in interactive systems.

\section{\bench}
\label{sec:benchmark}

\subsection{Task Definition}
\label{sec:task}

A game environment is defined as a tuple $\mathcal{E} = (\mathcal{S}, \mathcal{A}, T, s_0)$, where $\mathcal{S}$ denotes the state space, $\mathcal{A}$ the action space available to the agent, $T: \mathcal{S} \times \mathcal{A} \rightarrow \mathcal{S}$ the state transition function, and $s_0 \in \mathcal{S}$ the initial state.
Optionally, a documentation context $\mathcal{D}$ containing design documents and source code produced during game construction may be provided to the agent.
At each time step $t$, the agent observes state $s_t$, selects an action $a_t \in \mathcal{A}$, and the environment transitions to $s_{t+1} = T(s_t, a_t)$.
The agent interacts with the environment over multiple turns, forming an exploration trajectory $\tau = (s_0, a_0, s_1, a_1, \ldots, s_N)$.

Let $\mathcal{B} = \{B_1, B_2, \ldots, B_M\}$ denote the set of ground-truth bugs present in the environment.
After exploring $\mathcal{E}$, the agent produces a set of bug reports $\mathcal{R} = \{R_1, R_2, \ldots, R_K\}$, where each report $R_i$ contains a natural language description of the observed anomaly along with steps to reproduce it.
The objective of the agent is to maximize the coverage of $\mathcal{B}$ by $\mathcal{R}$, so that every bug in the environment is detected and described in sufficient detail for a software engineer to reproduce and fix it.
The general procedure is summarized in Algorithm~\ref{alg:bug_discovery}, and we define the formal evaluation protocol in Section~\ref{sec:evaluation}.

When $\mathcal{D} = \varnothing$, the agent operates in \emph{Player Exploring Mode}, relying solely on interactive observations to discover bugs from a player's perspective.
When $\mathcal{D}$ is provided, the agent operates in \emph{Quality Assurance Mode}, leveraging design specifications and source code to perform informed, specification-driven testing.
We evaluate both modes in Section~\ref{sec:experiments}.

\begin{algorithm}[t]
\caption{Task Definition of Quality Assurance Agent}
\label{alg:bug_discovery}
\begin{algorithmic}[1]
\Require Game environment $\mathcal{E} = (\mathcal{S}, \mathcal{A}, T, s_0)$, max steps $N$, optional documentation $\mathcal{D}$
\Ensure Bug report set $\mathcal{R}$
\State $s \gets s_0$, \quad $\mathcal{R} \gets \varnothing$, \quad $\tau \gets \varnothing$
\For{$t = 0, 1, \ldots, N$}
    \State $o_t \gets \textsc{Observe}(s_t)$
    \State $a_t \gets \textsc{Plan}(o_t, \tau, \mathcal{R}, \mathcal{D})$
    \State $s_{t+1}, r_t \gets T(s_t, a_t)$
    \State $o_{t+1} \gets \textsc{Observe}(s_{t+1})$
    \State $\tau \gets \tau \cup \{(o_t, a_t, o_{t+1})\}$
    
    \State $\hat{o}_{t+1} \gets \textsc{PredictExpectation}(o_t, a_t, \mathcal{D})$
    \State $\delta_t \gets \textsc{Reflect}(\hat{o}_{t+1}, o_{t+1})$
    
    \If{$\textsc{IsAnomaly}(\delta_t)$}
        \State $R \gets \textsc{GenerateReport}(\tau, \delta_t, \mathcal{D})$
        \State $\mathcal{R} \gets \mathcal{R} \cup \{R\}$
    \EndIf
\EndFor
\State \Return $\mathcal{R}$
\end{algorithmic}
\end{algorithm}

\subsection{Game Environment Builder}
\label{sec:builder}

To support scalable and controllable benchmark construction, all environments in \bench are developed by a hierarchical multi-agent collaboration system, which includes a Producer Agent and several working teams that simulates a professional game studio.
The Producer Agent decomposes high-level game concepts into structured proposal and distributes it to specialized teams responsible for design, programming, and art asset production.
Within each team, a Team Lead Agent further decomposes tasks based on dependencies and priorities, assigning subtasks to worker agents and coordinating progress. All agents share a support platform with reusable skills. Following the Agent Skills paradigm~\citep{zhang2025agentskills}, each skill is organized as a self-contained module with structured instructions and executable tools, enabling agents to discover and load capabilities on demand.
This multi-agent framework ensures structural coherence across design specifications, asset production, and code implementation. The overall architecture and more detailed operational principles are provided in Appendix~\ref{appendix:builder}.

All game environments are deployed as lightweight web applications.
To ensure a unified, agent-friendly interface, we adopt a strict frontend-backend separation architecture.
The backend encapsulates the core gameplay logic, exposes API endpoints for interaction, and handles state transitions triggered by agent actions.
The frontend renders game state updates received from the backend for human players and provides a standard interface for manual playtesting.
When a QA agent interacts with a game, it operates exclusively through the backend endpoints, which serve as callable tools to construct its action space. This design ensures that the observation space of the agent, including both game state and available actions, is semantically equivalent to what a human tester perceives through the frontend interface.

To prevent trivially simple environments, we introduce an iterative complexity scaling mechanism.
After an initial version of a game is generated, a QA agent performs a preliminary testing pass to estimate bug discoverability.
If the detected bug count falls below a predefined threshold $\tau$, the system automatically introduces additional gameplay features, mechanical interactions, or narrative branches to increase structural complexity.
This process iterates until the bug count meets or exceeds $\tau$.
Concurrently, each game is guaranteed to contain at least one fully functional gameplay trajectory, ensuring ecological validity and preventing unsolvable or broken states.

\subsection{Benchmark}
\label{sec:benchmark_data}

We instantiate the builder to construct \bench, a benchmark consisting of 30 diverse game environments and a total of 124 human-verified bugs spanning six core gameplay genres: Action, Adventure, Role-Playing, Strategy, Simulation, and Puzzle. Further statistics and game examples are provided in Appendix~\ref{appendix:environments}.

\paragraph{Discovery Difficulty.}
To provide a more detailed analysis, We define a three-level taxonomy for discovery difficulty based on the cognitive and reasoning demands required to detect a specific bug.
\begin{itemize}[leftmargin=*,labelsep=2mm]
    \item \diffeasy{Easy} bugs are surface-level perception inconsistencies 
          that can be identified from a single observation without multi-step reasoning.
    \item \diffmedium{Medium} bugs involve violations of gameplay logic or rule constraints 
          requiring the agent to reason about preconditions, action effects, and expected system behavior over short interaction sequences.
    \item \diffhard{Hard} bugs demand long-horizon consistency tracking across extended trajectories, 
          where contradictions only emerge when the agent integrates information over temporally separated states.
\end{itemize}

This taxonomy forms a structured progression from perceptual validation to rule-based reasoning and finally to long-horizon temporal consistency tracking. As shown in Figure~\ref{fig:benchmark}.
The benchmark exhibits a balanced structure centered around medium difficulty, while retaining meaningful proportions of both surface-level and long-horizon defects.

\paragraph{Ground-Truth Curation.}
The ground-truth bug dataset is established through a rigorous two-phase protocol that integrates automated discovery with expert validation.
In the initial phase, bug reports generated during the complexity scaling process (Section~\ref{sec:builder}) serve as preliminary candidates.
Subsequently, three professional QA engineers independently validate these candidates within each game environment, filtering out false positives and annotating confirmed bugs with structured metadata such as difficulty levels and reproduction steps.
Disagreements are resolved through majority voting to ensure annotation reliability. The labeling instructions can be found in Appendix~\ref{appendix:labeling}.




\subsection{Evaluation Metrics}
\label{sec:evaluation}

We evaluate the set of bug reports $\mathcal{R}$ produced by the agent against the ground-truth bug set $\mathcal{B}$.
A critic agent $f: \mathcal{R} \times \mathcal{B} \rightarrow \{0, 1\}$ determines whether a report $R_i$ successfully identifies a ground-truth bug $B_j$, based on the semantic correspondence between the description in $R_i$ and the annotation of $B_j$.
We define the set of successfully detected bugs as $\mathcal{B}^{+} = \{B_j \in \mathcal{B} \mid \exists\, R_i \in \mathcal{R},\; f(R_i, B_j) = 1\}$.
The primary evaluation metric is \textbf{Recall}, defined as 
\begin{equation}
\text{Recall} = \frac{|\mathcal{B}^{+}|}{|\mathcal{B}|}.
\end{equation}
We prioritize recall because the central objective of autonomous bug discovery is to maximize defect coverage. In practical QA workflows, false negatives carry substantially higher costs than false positives, as undetected defects may persist into production whereas spurious reports can be efficiently filtered by human reviewers.

\section{Baseline Agent}
\label{sec:agent}

We propose a baseline agent architecture that equips LLMs with dynamic exploration, reflective reasoning, feedback grounding, and memory management to support autonomous bug discovery over extended gameplay sessions.

\subsection{ReAct-Driven Exploration with Verification-Based Reflection}
\label{sec:react}

The agent follows the ReAct paradigm~\citep{yao2023react}, interleaving explicit reasoning with environment actions.
At each step $t$, given an observation $o_t$, the agent generates reasoning traces regarding the current state and expected outcomes, selects an action $a_t$ from the available tool set, and transitions to the subsequent observation $o_{t+1}$.

To enhance sensitivity to anomalies, we augment standard ReAct with a step-level reflection and verification mechanism.
After each transition $(o_t, a_t, o_{t+1})$, the agent critically evaluates whether the observed outcome aligns with its internal expectation of correct game behavior.

Upon detecting a discrepancy, the agent formulates a preliminary bug hypothesis consisting of 
(i) the triggering action,
(ii) observed behavior,
(iii) expected behavior,
and (iv) potential violation type.
Rather than immediately reporting, the agent initiates a local verification phase to collect corroborating evidence through targeted reproduction attempts.
Based on reproducibility and deviation magnitude, a confidence score is assigned, with only candidates exceeding the threshold are promoted to final bug reports, thereby mitigating false positives.
This mechanism transforms the agent's role from a passive trajectory generator to an active behavioral verifier, tightly aligning its reasoning process with the objective of autonomous bug discovery.

\subsection{Hierarchical Memory Module}
\label{sec:memory}

To overcome the context-window limitations of LLMs in long-horizon bug discovery, we introduce a hierarchical memory architecture that separates short-term trajectory tracking from long-term experiential accumulation.

\textbf{In-Session Memory.} Within a single gameplay session, the agent maintains a structured working memory that tracks the evolution of the game state. As interaction histories grow, earlier trajectory segments are periodically compressed using a summarization module. These summaries retain semantically critical information, including visited locations, acquired items, triggered events, unresolved anomalies, and tentative bug hypotheses.

To balance fidelity and scalability, we adopt a sliding-window strategy where the most recent $k$ interaction steps are preserved in full detail, while older steps are replaced by compact state summaries. This design enables long-horizon reasoning while remaining within the model's context constraints. Importantly, the summarization process is not purely extractive but abstraction-oriented, as it preserves causal structure (e.g., ``after picking up item X, event Y becomes available") rather than raw textual logs. This abstraction supports reasoning about delayed effects and multi-step inconsistencies.

\textbf{Cross-Session Memory.} Thorough QA tasks frequently require restarting and re-exploring a game from different initial conditions. To mirror this realistic testing workflow, we maintain a persistent cross-session memory store for each game.

After each session, the agent distills its accumulated experience into a structured summary that captures explored regions, confirmed bugs, unresolved hypotheses, unexplored branches, and priority testing targets. This summary is injected into the initial context of subsequent sessions. By separating intra-session trajectory management from inter-session knowledge accumulation, the agent progressively builds a coherent testing strategy across multiple restarts. This hierarchical memory design improves exploration efficiency, reduces redundant coverage, and encourages systematic testing rather than random wandering.

\section{Experiments}
\label{sec:experiments}

\subsection{Experimental Setup}
\label{sec:setup}

\textbf{Models.}
We evaluate a diverse suite of frontier LLMs spanning open-source and closed-source families, including instruct and thinking variants. All models use officially recommended decoding parameters; otherwise, we adopt greedy sampling as the default strategy.

\textbf{Settings.}
Each model serves as the backbone of the baseline agent described in Section~\ref{sec:agent}. As defined in Section~\ref{sec:task}, we evaluate each model under both Player Exploring Mode and Quality Assurance Mode. For each game in \bench, the agent is given a maximum budget of $T$ interaction steps.
We evaluate across four step budgets ($T \in \{50, 100, 200, 500\}$) under both modes to examine how the extent of exploration affects bug detection coverage.

\textbf{Metrics.}
We adopt Recall as the primary metric, computed via automated evaluation by critic agent.

\subsection{Main Results}
\label{sec:main_results}
Following the setup above, we compare a wide range of mainstream LLMs. Table~\ref{tab:leaderboard} reports the performance of each model under both testing modes across all step budgets. The experiment results reveal several insights and patterns across modes and model families.

\begin{table*}[htbp]
 \centering
 \small
 \setlength{\tabcolsep}{4pt}

 \resizebox{\textwidth}{!}{%
 \begin{tabular}{@{}l cccc cccc c@{}}
 \toprule
 \multirow{2}{*}{\textbf{Model}} & \multicolumn{4}{c}{\textbf{Player Exploring Mode}} & \multicolumn{4}{c}{\textbf{Quality Assurance Mode}} & \multirow{2}{*}{\textbf{Best Performance}} \\
 \cmidrule(lr){2-5} \cmidrule(lr){6-9}
 & \textbf{50} & \textbf{100} & \textbf{200} & \textbf{500} & \textbf{50} & \textbf{100} & \textbf{200} & \textbf{500} & \\
 \midrule
 \multicolumn{10}{c}{\textit{LLMs in Instruct Mode}} \\
 \midrule
 \rowcolor{lightgray}
 Claude-4.6-Opus          & 14.52 & 20.97 & 25.81 & 31.45 & 22.58 & 28.23 & 31.45 & 37.90 & 37.90 \\
 Claude-4.5-Sonnet        & 11.29  & 16.13 & 18.55 & 20.97 & 17.74 & 25.00 & 28.23 & 32.26 & 32.26 \\
 \rowcolor{lightgray}
 GPT-5.2                  & 7.26  & 10.48 & 12.90 & 14.52 & 11.29 & 16.94 & 19.35 & 22.58 & 22.58 \\
 Kimi-K2.5-1T-A32B        & 6.45  & 9.68  & 11.29 & 13.71 & 10.48 & 15.32 & 17.74 & 20.97 & 20.97 \\
 \rowcolor{lightgray}
 Gemini-3-Flash           & 6.45  & 8.87  & 10.48 & 12.10 & 9.68  & 13.71 & 16.13 & 19.35 & 19.35 \\
 DeepSeek-V3.2            & 6.45  & 9.68  & 10.48 & 12.90 & 9.68  & 14.52 & 16.94 & 20.16 & 20.16 \\
 \rowcolor{lightgray}
 Llama-3.1-8B             & 2.42  & 3.23  & 4.84  & 5.65  & 4.03  & 5.65  & 7.26  & 8.87  & 8.87  \\
 Llama-3.1-70B            & 4.03  & 6.45  & 8.06  & 9.68  & 6.45  & 9.68  & 12.10 & 14.52 & 14.52 \\
 \rowcolor{lightgray}
 Qwen3-8B                 & 4.03  & 5.65  & 6.45  & 7.26  & 6.45  & 8.06  & 9.68  & 10.48 & 10.48 \\
 Qwen3-32B                & 4.84  & 7.26  & 9.68  & 10.48 & 6.45  & 11.29 & 14.52 & 15.32 & 15.32 \\
 \rowcolor{lightgray}
 Qwen3-235B-A22B          & 5.65  & 9.68  & 10.48 & 12.10 & 8.87  & 14.52 & 16.13 & 18.55 & 18.55 \\
 Qwen3.5-397B-A17B        & 8.06  & 11.29 & 13.71 & 15.32 & 12.10 & 17.74 & 20.97 & 24.19 & 24.19 \\
 \midrule
 \multicolumn{10}{c}{\textit{LLMs in Thinking Mode}} \\
 \midrule
\rowcolor{lightgray}
Claude-4.6-Opus-Thinking       & 16.94 & 23.39 & 29.03 & 35.48 & \textbf{25.00} & \textbf{34.68} & \textbf{41.13} & \textbf{48.39} & \textbf{48.39} \\
Claude-4.5-Sonnet-Thinking     & 12.10 & 17.74 & 21.77 & 26.61 & 19.35 & 25.81 & 30.65 & 37.10 & 37.10 \\
\rowcolor{lightgray}
OpenAI-o3                      & 11.29 & 16.13 & 20.97 & 25.00 & 17.74 & 25.00 & 29.84 & 34.68 & 34.68 \\
Kimi-K2.5-1T-A32B-Thinking     & 8.87  & 12.90 & 16.13 & 20.16 & 14.52 & 20.16 & 24.19 & 28.23 & 28.23 \\
\rowcolor{lightgray}
Gemini-3-Pro                   & 10.48 & 15.32 & 19.35 & 23.39 & 16.94 & 22.58 & 27.42 & 33.06 & 33.06 \\
DeepSeek-R1                    & 11.29 & 17.74 & 22.58 & 27.42 & 19.35 & 27.42 & 32.26 & 37.90 & 37.90 \\
\rowcolor{lightgray}
Qwen3-8B-Thinking              & 7.26  & 10.48 & 12.90 & 16.13 & 12.10 & 16.94 & 20.97 & 24.19 & 24.19 \\
Qwen3-32B-Thinking             & 9.68  & 14.52 & 19.35 & 24.19 & 15.32 & 23.39 & 29.03 & 33.87 & 33.87 \\
\rowcolor{lightgray}
Qwen3-235B-A22B-Thinking       & 10.48 & 16.13 & 20.97 & 25.00 & 18.55 & 25.00 & 30.65 & 35.48 & 35.48 \\
Qwen3.5-397B-A17B-Thinking     & 13.71 & 19.35 & 25.00 & 30.65 & 20.97 & 28.23 & 35.48 & 41.13 & 41.13 \\
 \bottomrule
 \end{tabular}
 }
 \caption{\bench Leaderboard. We report Recall (\%) under two testing modes across four step budgets. \textbf{Bold} values indicate the highest score. Best Performance denotes the highest score achieved by each model across all settings.}
\label{tab:leaderboard}
\end{table*}

\textbf{Challenging Benchmark.}
Autonomous bug discovery remains highly challenging for all evaluated models.
Even the best-performing configuration, Claude-4.6-Opus under Quality Assurance Mode with 500 steps, achieves only 48.39\%, leaving over half of the bugs undetected.
This confirms that bug discovery constitutes a substantially harder capability than general code generation or issue resolution, where frontier models routinely exceed 70\% on comparable benchmarks such as SWE-Bench Verified~\citep{chowdhury2024swebenchverified}. A detailed comparison of frontier model performance on SWE-Bench Verified versus GBQA is provided in Appendix~\ref{app:model-comparison}.

\textbf{Scaling Law.}
While standard scaling trends persist in this setting, as evidenced by consistent performance gains with model size (e.g., the Qwen3 series), reasoning capability proves to be more parameter-efficient than merely increasing model scale. For instance, Qwen3-32B-Thinking (33.87\%) significantly outperforms the much larger Llama-3.1-70B (14.52\%) and even rivals the massive Qwen3-235B-A22B (18.55\%). This suggests that for bug discovery, which demands sustained multi-step reasoning and dynamic state verification, inference-time scaling is more critical than parameter scaling alone.

\textbf{Testing Mode.} The Quality Assurance mode consistently outperforms the Player Exploring mode across all evaluated models and step budgets. Access to design artifacts and source code enables specification-driven testing, allowing agents to establish precise behavioral expectations and, consequently, detect finer-grained violations. Nevertheless, even with comprehensive documentation, performance remains substantially suboptimal. This persistent gap indicates that the primary bottleneck lies not in context information scarcity, but in two inherent limitations of current LLMs: 
(i) susceptibility to hallucinations and logical inconsistencies during complex multi-step reasoning, coupled with error accumulation and state-tracking ambiguity in long-horizon tasks; and 
(ii) a pronounced deficit in systematic testing heuristics, attributable to the scarcity of QA-specific RL training. Consequently, these models lack the structured, efficient, and hypothesis-driven exploration strategies routinely employed by experienced QA engineers.

\subsection{Case Study}
\label{sec:case_study}
To demonstrate the practical utility of \bench, we conduct a case study on a fully autonomous detection-to-patch pipeline. Additional experimental details and results are provided in Appendix~\ref{appendix:case-study}.

\subsection{Reliability of \bench}
\label{sec:reliability}

\begin{figure}[t]
\centering
\begin{minipage}[c]{0.48\linewidth}
    \centering

    \resizebox{\linewidth}{!}{
        \begin{tabular}{@{}lcc@{}}
        \toprule
        \textbf{Annotation Set} & \textbf{Count} & \textbf{Krippendorff's $\alpha$ [95\% CI]} \\
        \midrule
        Valid Bug        & 124 & $0.8920\,[-0.0613, +0.0614]$ \\
        Non-Bug          & 254 & $0.9180\,[-0.0462, +0.0461]$ \\
        \midrule
        \textbf{Overall Candidates} & 378 & $\mathbf{0.9010}\,\mathbf{[-0.0391, +0.0389]}$ \\
        \bottomrule
        \end{tabular}
    }
    \captionof{table}{Inter-Annotator Agreement analysis for human annotation in bug classification.}
    \label{tab:iaa_human}

    \vspace{0.5cm}

    \resizebox{\linewidth}{!}{
        \centering
        \begin{tabular}{lcc}
        \toprule
        \textbf{Model} & \textbf{Pearson $\rho$ [95\% CI]} & \textbf{p-value} \\
        \midrule
        Gemini-3-Pro      & 0.858 [$-0.0548$, $0.0404$]          & $< 0.0001$ \\
        Claude-4.6-Opus   & 0.821 [$-0.0672$, $0.0502$]          & $< 0.0001$ \\
        DeepSeek-R1       & 0.807 [$-0.0717$, $0.0538$]          & $< 0.0001$ \\
        GPT-5.2           & $\mathbf{0.903\ [-0.0273, 0.0196]}$  & $< 0.0001$ \\
        \bottomrule
        \end{tabular}
    }
    \captionof{table}{Pearson correlation coefficients and p-values of different models and human evaluators.}
    \label{tab:critic}
\end{minipage}
\hfill
\begin{minipage}[c]{0.50\linewidth}
    \centering
    \includegraphics[width=0.85\textwidth]{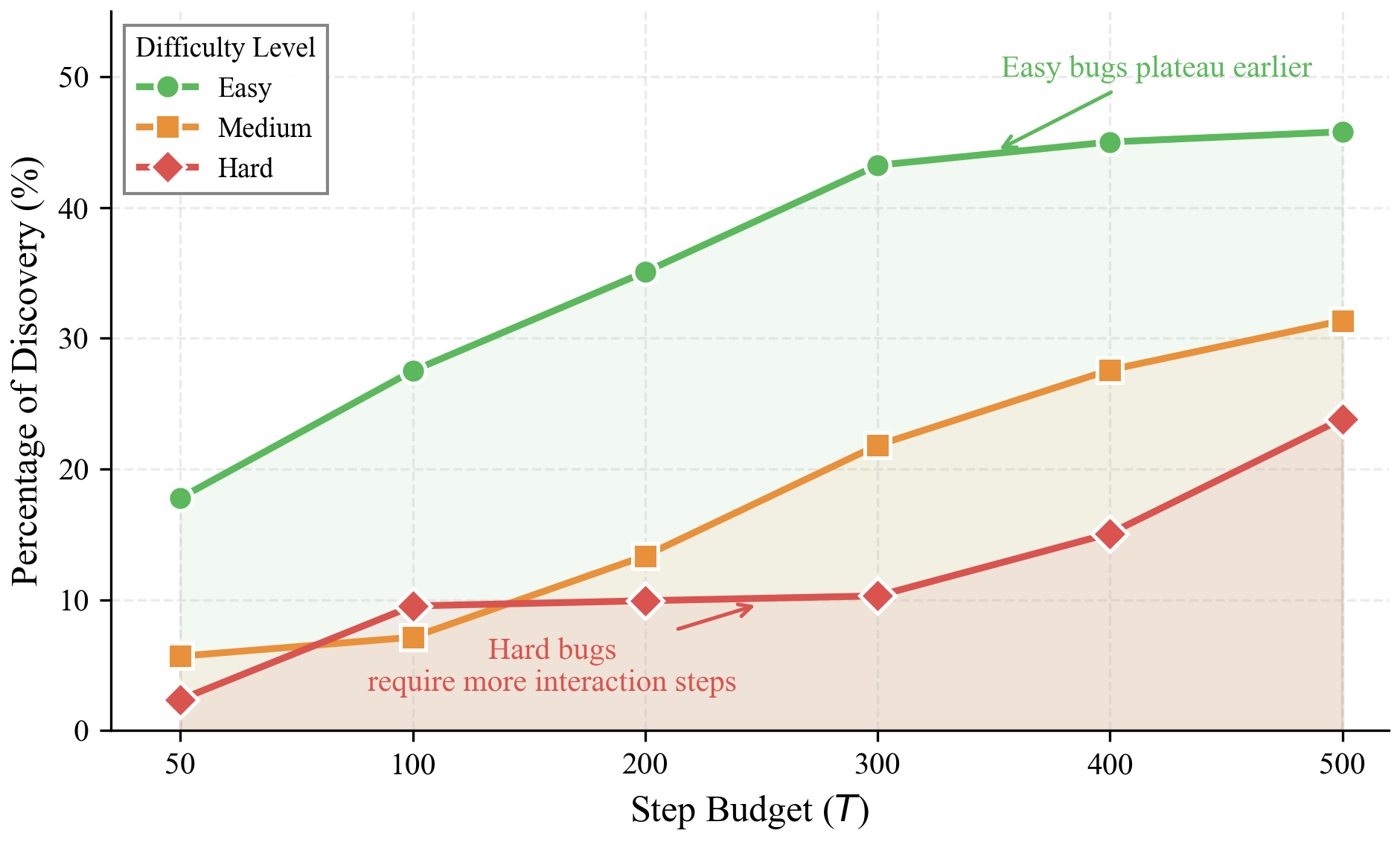}
    \caption{Percentage of bug discovery by difficulty level across step budgets. Easy bugs are largely discovered within the first 300 steps, while hard bugs require substantially more interaction steps and remain growing even at nearly 500 steps.}
    \label{fig:steps}
\end{minipage}
\end{figure}

\textbf{IAA Analysis for Benchmark Annotation.}
To quantify the reliability of the bug annotations in \bench, we conduct an Inter-Annotator Agreement (IAA) analysis using Krippendorff's $\alpha$~\citep{krippendorff2018content}.
As shown in Table~\ref{tab:iaa_human}, three annotators independently label each of the 378 candidate annotations as either a valid bug or a non-bug.
The dataset achieves an overall $\alpha$ of 0.901, indicating that the labeling instructions successfully standardize expert judgments despite the inherent subjectivity of bug characterization.

\textbf{Critic Agent as Evaluator.}
To further validate the automated evaluation pipeline, we measure its agreement with human ratings using Pearson correlation coefficient~\citep{pearson1901liii} on a held-out validation set. As reported in Table~\ref{tab:critic}, all four backbone LLMs achieve high correlations, confirming that the Critic Agent serves as a reliable proxy for human evaluation.
GPT-5.2 achieves the highest correlation ($\rho = 0.903$) and is therefore adopted as the default backbone for all reported results.

\subsection{Ablation Studies}
\label{sec:ablation}


We conduct ablation experiments using Claude-4.6-Opus under Quality Assurance Mode to isolate the contributions of individual architectural components.

\textbf{Step Budget Analysis.}
We vary the step budget $T$ to study the trade-off between computational cost and bug discovery, stratified by difficulty level.
As shown in Figure~\ref{fig:steps}, \diffeasy{Easy} bugs are largely discovered within the first 300 steps, while \diffmedium{Medium} bugs follow a similar but lower trajectory, reaching about 30\% at 500 steps.
\diffhard{Hard} bugs show the strongest dependence on step budget, with no clear saturation trend.
This pattern reveals that Easy bugs require perceptual checking, Medium bugs short-horizon rule inference, while Hard bugs sustained state tracking over long interactions.

\begin{figure}[t]
\centering
    \includegraphics[width=\linewidth]{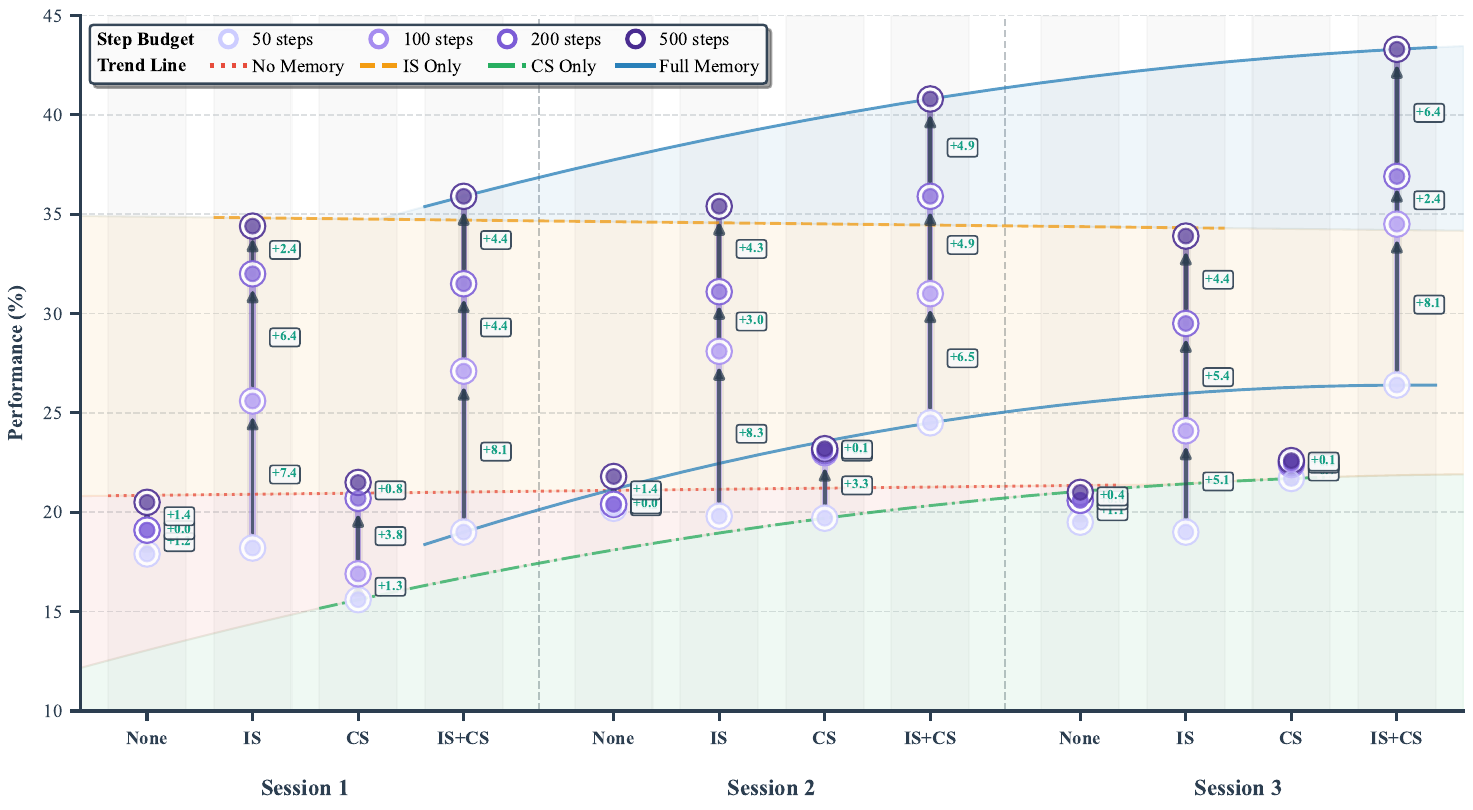}
    \caption{
    Ablation study of memory module.
    Each cluster corresponds to a session, and vertical arrows indicate performance gains as the step budget increases. The four trend lines illustrate the aggregated trend for same memory settings across sessions.
    }
    \label{fig:memory-ablation}
\end{figure}

\textbf{Memory Ablation.}
As illustrated in Figure~\ref{fig:memory-ablation}, there are four experimental configurations, namely no memory, in-session memory (IS), cross-session memory (CS), and the full memory module (IS+CS).
Without memory, the agent frequently revisits tested states, causing early recall saturation.
Although IS memory eliminates within-session loops, it necessitates re-exploration across sessions. Conversely, CS memory enables warm-start exploration but fails to mitigate in-session redundancy.
The full memory module integrates strategic initialization across sessions with loop prevention within sessions. Consequently, its performance trend line consistently dominates other memory settings and exhibits clear gains across sessions at equivalent step budgets, demonstrating complementary benefits from intra-session trajectory tracking and inter-session knowledge accumulation.

\section{Conclusion}
\label{sec:conclusion}

We presented \bench, a scalable benchmark for evaluating the autonomous bug discovery capabilities of LLMs in interactive game environments.
Our experimental results reveal that, despite strong performance in code generation and repair tasks, state-of-the-art LLMs remain substantially limited in bug discovery, particularly for long-horizon and state-dependent errors.
These findings highlight a significant gap between current agent capabilities and the real-world demands of quality assurance. By providing standardized environments, quantitative metrics, and reliable evaluation, \bench offers a foundation for the principled design and comparison of future QA agents. We believe this benchmark opens a new research direction at the intersection of agentic reasoning and software development. In future work, we will extend \bench beyond games towards broader domains, incorporating multimodal perception and GUI interaction to better reflect real-world scenarios.

\bibliography{main}

@inproceedings{wang2023voyager,
  title={Voyager: An Open-Ended Embodied Agent with Large Language Models},
  author={Wang, Guanzhi and Xie, Yuqi and Jiang, Yunfan and Mandlekar, Ajay and Xiao, Chaowei and Zhu, Yuke and Fan, Linxi and Anandkumar, Anima},
  booktitle={NeurIPS},
  year={2023}
}

@inproceedings{fan2022minedojo,
  title={MineDojo: Building Open-Ended Embodied Agents with Internet-Scale Knowledge},
  author={Fan, Linxi and Wang, Guanzhi and Jiang, Yunfan and Mandlekar, Ajay and Yang, Yuncong and Zhu, Haoyi and Tang, Andrew and Huang, De-An and Zhu, Yuke and Anandkumar, Anima},
  booktitle={NeurIPS},
  year={2022}
}

@article{park2023generative,
  title={Generative Agents: Interactive Simulacra of Human Behavior},
  author={Park, Joon Sung and O'Brien, Joseph C and Cai, Carrie J and Morris, Meredith Ringel and Liang, Percy and Bernstein, Michael S},
  journal={arXiv preprint arXiv:2304.03442},
  year={2023}
}

@misc{tan2024cradle,
      title={Cradle: Empowering Foundation Agents Towards General Computer Control}, 
      author={Weihao Tan and Wentao Zhang and Xinrun Xu and Haochong Xia and Ziluo Ding and Boyu Li and Bohan Zhou and Junpeng Yue and Jiechuan Jiang and Yewen Li and Ruyi An and Molei Qin and Chuqiao Zong and Longtao Zheng and Yujie Wu and Xiaoqiang Chai and Yifei Bi and Tianbao Xie and Pengjie Gu and Xiyun Li and Ceyao Zhang and Long Tian and Chaojie Wang and Xinrun Wang and Börje F. Karlsson and Bo An and Shuicheng Yan and Zongqing Lu},
      year={2024},
      eprint={2403.03186},
      archivePrefix={arXiv},
      primaryClass={cs.AI},
      url={https://arxiv.org/abs/2403.03186}, 
}

@inproceedings{zhou2024webarena,
    title={WebArena: A Realistic Web Environment for Building Autonomous Agents},
    author={Shuyan Zhou and Frank F. Xu and Hao Zhu and Xuhui Zhou and Robert Lo and Abishek Sridhar and Xianyi Cheng and Tianyue Ou and Yonatan Bisk and Daniel Fried and Uri Alon and Graham Neubig},
    booktitle={The Twelfth International Conference on Learning Representations},
    year={2024},
    url={https://openreview.net/forum?id=oKn9c6ytLx}
}

@inproceedings{liu2024agentbench,
    title={AgentBench: Evaluating {LLM}s as Agents},
    author={Xiao Liu and Hao Yu and Hanchen Zhang and Yifan Xu and Xuanyu Lei and Hanyu Lai and Yu Gu and Hangliang Ding and Kaiwen Men and Kejuan Yang and Shudan Zhang and Xiang Deng and Aohan Zeng and Zhengxiao Du and Chenhui Zhang and Sheng Shen and Tianjun Zhang and Yu Su and Huan Sun and Minlie Huang and Yuxiao Dong and Jie Tang},
    booktitle={The Twelfth International Conference on Learning Representations},
    year={2024},
    url={https://openreview.net/forum?id=zAdUB0aCTQ}
}

@inproceedings{jimenez2024swebench,
  title={SWE-bench: Can Language Models Resolve Real-world GitHub Issues?},
  author={Jimenez, Carlos E. and Yang, John and Wettig, Alexander and Yao, Shunyu and Pei, Kexin and Press, Ofir and Narasimhan, Karthik},
  booktitle={The Twelfth International Conference on Learning Representations (ICLR)},
  year={2024},
  url={https://openreview.net/forum?id=VTF8yNQM66}
}

@misc{chowdhury2024swebenchverified,
  author = {Neil Chowdhury and James Aung and Chan Jun Shern and Oliver Jaffe and Dane Sherburn and Giulio Starace and Evan Mays and Rachel Dias and Marwan Aljubeh and Mia Glaese and Carlos E. Jimenez and John Yang and Leyton Ho and Tejal Patwardhan and Kevin Liu and Aleksander Madry},
  title = {Introducing SWE-bench Verified},
  year = {2024},
  month = aug,
  url = {https://openai.com/index/introducing-swe-bench-verified/},
}

@misc{aleithan2024swebenchenhancedcodingbenchmark,
      title={SWE-Bench+: Enhanced Coding Benchmark for LLMs}, 
      author={Reem Aleithan and Haoran Xue and Mohammad Mahdi Mohajer and Elijah Nnorom and Gias Uddin and Song Wang},
      year={2024},
      eprint={2410.06992},
      archivePrefix={arXiv},
      primaryClass={cs.SE},
      url={https://arxiv.org/abs/2410.06992}, 
}

@inproceedings{yao2023react,
  title={ReAct: Synergizing Reasoning and Acting in Language Models},
  author={Yao, Shunyu and Zhao, Jeffrey and Yu, Dian and Du, Nan and Shafran, Izhak and Narasimhan, Karthik and Cao, Yuan},
  booktitle={ICLR},
  year={2023}
}

@inproceedings{qian2024chatdev,
    title = "{C}hat{D}ev: Communicative Agents for Software Development",
    author = "Qian, Chen  and
      Liu, Wei  and
      Liu, Hongzhang  and
      Chen, Nuo  and
      Dang, Yufan  and
      Li, Jiahao  and
      Yang, Cheng  and
      Chen, Weize  and
      Su, Yusheng  and
      Cong, Xin  and
      Xu, Juyuan  and
      Li, Dahai  and
      Liu, Zhiyuan  and
      Sun, Maosong",
    editor = "Ku, Lun-Wei  and
      Martins, Andre  and
      Srikumar, Vivek",
    booktitle = "Proceedings of the 62nd Annual Meeting of the Association for Computational Linguistics (Volume 1: Long Papers)",
    month = aug,
    year = "2024",
    address = "Bangkok, Thailand",
    publisher = "Association for Computational Linguistics",
    url = "https://aclanthology.org/2024.acl-long.810/",
    doi = "10.18653/v1/2024.acl-long.810",
    pages = "15174--15186"
}

@inproceedings{hong2023metagpt,
  title={MetaGPT: Meta programming for a multi-agent collaborative framework},
  author={Hong, Sirui and Zhuge, Mingchen and Chen, Jonathan and Zheng, Xiawu and Cheng, Yuheng and Wang, Jinlin and Zhang, Ceyao and Wang, Zili and Yau, Steven Ka Shing and Lin, Zijuan and others},
  booktitle={The twelfth international conference on learning representations},
  year={2024}
}

@article{xia2024agentless,
  title={Agentless: Demystifying LLM-based Software Engineering Agents},
  author={Xia, Chunqiu Steven and Deng, Yinlin and Dunn, Soren and Zhang, Lingming},
  journal={arXiv preprint arXiv:2407.01489},
  year={2024}
}

@article{titan2025,
  title={Leveraging {LLM} Agents for Automated Video Game Testing},
  author={Wang, Chengjia and Tang, Lanling and Yuan, Ming and Yu, Jiongchi and Xie, Xiaofei and Bu, Jiajun},
  journal={arXiv preprint arXiv:2509.22170},
  year={2025}
}

@article{orak2025,
  title={Orak: A Foundational Benchmark for Training and Evaluating {LLM} Agents on Diverse Video Games},
  author={Park, Dongmin and Kim, Minkyu and Choi, Beongjun and Kim, Junhyuck and Lee, Keon and Lee, Jonghyun and Park, Inkyu and Lee, Byeong-Uk and Hwang, Jaeyoung and Ahn, Jaewoo and Mahabaleshwarkar, Ameya S. and Kartal, Bilal and Biswas, Pritam and Suhara, Yoshi and Lee, Kangwook and Cho, Jaewoong},
  journal={arXiv preprint arXiv:2506.03610},
  year={2025}
}

@misc{mu2025synergizingcodecoveragegameplay,
      title={Synergizing Code Coverage and Gameplay Intent: Coverage-Aware Game Playtesting with LLM-Guided Reinforcement Learning}, 
      author={Enhong Mu and Minami Yoda and Yan Zhang and Mingyue Zhang and Yutaka Matsuno and Jialong Li},
      year={2025},
      eprint={2512.12706},
      archivePrefix={arXiv},
      primaryClass={cs.AI},
      url={https://arxiv.org/abs/2512.12706}, 
}

@inproceedings{guo2025repoaudit,
    title={RepoAudit: An Autonomous {LLM}-Agent for Repository-Level Code Auditing},
    author={Jinyao Guo and Chengpeng Wang and Xiangzhe Xu and Zian Su and Xiangyu Zhang},
    booktitle={Forty-second International Conference on Machine Learning},
    year={2025},
    url={https://openreview.net/forum?id=TXcifVbFpG}
}

@article{bugstone2025,
  title={One Bug, Hundreds Behind: {LLMs} for Large-Scale Bug Discovery},
  author={Wu, Qiushi and Xiao, Yue and Kirat, Dhilung and Eykholt, Kevin and Jang, Jiyong and Schales, Douglas Lee},
  journal={arXiv preprint arXiv:2510.14036},
  year={2025}
}

@article{pearson1901liii,
  title={LIII. On lines and planes of closest fit to systems of points in space},
  author={Pearson, Karl},
  journal={The London, Edinburgh, and Dublin philosophical magazine and journal of science},
  volume={2},
  number={11},
  pages={559--572},
  year={1901},
  publisher={Taylor \& Francis}
}

@book{krippendorff2018content,
  title={Content Analysis: An Introduction to Its Methodology},
  author={Krippendorff, Klaus},
  year={2018},
  edition={4th},
  publisher={SAGE Publications}
}

@misc{anthropic2025claudecode,
  title = {Claude Code},
  author = {{Anthropic}},
  year = {2025},
  note = {\url{https://claude.com/product/claude-code}},
}

@misc{cursor2024editor,
  title = {Cursor},
  author = {{Anysphere}},
  year = {2024},
  note = {\url{https://cursor.com/product}},
}

@misc{openai2025codex,
  title = {OpenAI Codex},
  author = {{OpenAI}},
  year = {2025},
  note = {\url{https://openai.com/codex/}},
}

@misc{karpathy2025vibecoding,
  title = {Concept of Vibe Coding},
  author = {Karpathy, Andrej},
  year = {2025},
  month = feb,
  note = {\url{https://x.com/karpathy/status/1886192184808149383}},
  howpublished = {X Post}
}

@inproceedings{Myers1979ArtOS,
  title={Art of Software Testing},
  author={Glenford J. Myers},
  year={1979},
  url={https://api.semanticscholar.org/CorpusID:59854592}
}

@book{ammann2016introduction,
  title = {Introduction to Software Testing},
  author = {Ammann, Paul and Offutt, Jeff},
  year = {2016},
  place={Cambridge},
  publisher = {Cambridge University Press},
  edition = {2nd}
}

@misc{zhang2025agentskills,
  title = {Equipping Agents for the Real World with Agent Skills},
  author = {Zhang, Barry and Lazuka, Keith and Murag, Mahesh},
  year = {2025},
  month = oct,
  note = {\url{https://claude.com/blog/equipping-agents-for-the-real-world-with-agent-skills}},
  organization = {Anthropic}
}

@misc{anthropic2026opus46card,
  title={Claude Opus 4.6 System Card},
  author={Anthropic},
  year={2026},
  month = feb,
  note={\url{https://www.anthropic.com/claude-opus-4-6-system-card}},
}

@misc{openai2025gpt52,
  title={Update to GPT-5 System Card: GPT-5.2},
  author={OpenAI},
  year={2025},
  month = dec,
  note={\url{https://openai.com/index/gpt-5-system-card-update-gpt-5-2/}},
}

@misc{cao2026qwen3codernexttechnicalreport,
      title={Qwen3-Coder-Next Technical Report}, 
      author={Ruisheng Cao and Mouxiang Chen and Jiawei Chen and Zeyu Cui and Yunlong Feng and Binyuan Hui and Yuheng Jing and Kaixin Li and Mingze Li and Junyang Lin and Zeyao Ma and Kashun Shum and Xuwu Wang and Jinxi Wei and Jiaxi Yang and Jiajun Zhang and Lei Zhang and Zongmeng Zhang and Wenting Zhao and Fan Zhou},
      year={2026},
      eprint={2603.00729},
      archivePrefix={arXiv},
      primaryClass={cs.CL},
      url={https://arxiv.org/abs/2603.00729}, 
}

@misc{google2026gemini31pro,
  title        = {Gemini 3.1 Pro Model Card},
  author       = {{Google DeepMind}},
  year         = {2026},
  month        = feb,
  url          = {https://deepmind.google/models/model-cards/gemini-3-1-pro/},
}

@misc{anthropic2025claudesonnet45,
  title = {Introducing Claude Sonnet 4.5},
  author = {{Anthropic}},
  year = {2025},
  month = sep,
  url = {https://www.anthropic.com/news/claude-sonnet-4-5},
}

@article{Guo_2025,
   title={DeepSeek-R1 incentivizes reasoning in LLMs through reinforcement learning},
   volume={645},
   ISSN={1476-4687},
   url={http://dx.doi.org/10.1038/s41586-025-09422-z},
   DOI={10.1038/s41586-025-09422-z},
   number={8081},
   journal={Nature},
   publisher={Springer Science and Business Media LLC},
   author={Guo, Daya and Yang, Dejian and Zhang, Haowei and Song, Junxiao and Wang, Peiyi and Zhu, Qihao and Xu, Runxin and Zhang, Ruoyu and Ma, Shirong and Bi, Xiao and Zhang, Xiaokang and Yu, Xingkai and Wu, Yu and Wu, Z. F. and Gou, Zhibin and Shao, Zhihong and Li, Zhuoshu and Gao, Ziyi and Liu, Aixin and Xue, Bing and Wang, Bingxuan and Wu, Bochao and Feng, Bei and Lu, Chengda and Zhao, Chenggang and Deng, Chengqi and Ruan, Chong and Dai, Damai and Chen, Deli and Ji, Dongjie and Li, Erhang and Lin, Fangyun and Dai, Fucong and Luo, Fuli and Hao, Guangbo and Chen, Guanting and Li, Guowei and Zhang, H. and Xu, Hanwei and Ding, Honghui and Gao, Huazuo and Qu, Hui and Li, Hui and Guo, Jianzhong and Li, Jiashi and Chen, Jingchang and Yuan, Jingyang and Tu, Jinhao and Qiu, Junjie and Li, Junlong and Cai, J. L. and Ni, Jiaqi and Liang, Jian and Chen, Jin and Dong, Kai and Hu, Kai and You, Kaichao and Gao, Kaige and Guan, Kang and Huang, Kexin and Yu, Kuai and Wang, Lean and Zhang, Lecong and Zhao, Liang and Wang, Litong and Zhang, Liyue and Xu, Lei and Xia, Leyi and Zhang, Mingchuan and Zhang, Minghua and Tang, Minghui and Zhou, Mingxu and Li, Meng and Wang, Miaojun and Li, Mingming and Tian, Ning and Huang, Panpan and Zhang, Peng and Wang, Qiancheng and Chen, Qinyu and Du, Qiushi and Ge, Ruiqi and Zhang, Ruisong and Pan, Ruizhe and Wang, Runji and Chen, R. J. and Jin, R. L. and Chen, Ruyi and Lu, Shanghao and Zhou, Shangyan and Chen, Shanhuang and Ye, Shengfeng and Wang, Shiyu and Yu, Shuiping and Zhou, Shunfeng and Pan, Shuting and Li, S. S. and Zhou, Shuang and Wu, Shaoqing and Yun, Tao and Pei, Tian and Sun, Tianyu and Wang, T. and Zeng, Wangding and Liu, Wen and Liang, Wenfeng and Gao, Wenjun and Yu, Wenqin and Zhang, Wentao and Xiao, W. L. and An, Wei and Liu, Xiaodong and Wang, Xiaohan and Chen, Xiaokang and Nie, Xiaotao and Cheng, Xin and Liu, Xin and Xie, Xin and Liu, Xingchao and Yang, Xinyu and Li, Xinyuan and Su, Xuecheng and Lin, Xuheng and Li, X. Q. and Jin, Xiangyue and Shen, Xiaojin and Chen, Xiaosha and Sun, Xiaowen and Wang, Xiaoxiang and Song, Xinnan and Zhou, Xinyi and Wang, Xianzu and Shan, Xinxia and Li, Y. K. and Wang, Y. Q. and Wei, Y. X. and Zhang, Yang and Xu, Yanhong and Li, Yao and Zhao, Yao and Sun, Yaofeng and Wang, Yaohui and Yu, Yi and Zhang, Yichao and Shi, Yifan and Xiong, Yiliang and He, Ying and Piao, Yishi and Wang, Yisong and Tan, Yixuan and Ma, Yiyang and Liu, Yiyuan and Guo, Yongqiang and Ou, Yuan and Wang, Yuduan and Gong, Yue and Zou, Yuheng and He, Yujia and Xiong, Yunfan and Luo, Yuxiang and You, Yuxiang and Liu, Yuxuan and Zhou, Yuyang and Zhu, Y. X. and Huang, Yanping and Li, Yaohui and Zheng, Yi and Zhu, Yuchen and Ma, Yunxian and Tang, Ying and Zha, Yukun and Yan, Yuting and Ren, Z. Z. and Ren, Zehui and Sha, Zhangli and Fu, Zhe and Xu, Zhean and Xie, Zhenda and Zhang, Zhengyan and Hao, Zhewen and Ma, Zhicheng and Yan, Zhigang and Wu, Zhiyu and Gu, Zihui and Zhu, Zijia and Liu, Zijun and Li, Zilin and Xie, Ziwei and Song, Ziyang and Pan, Zizheng and Huang, Zhen and Xu, Zhipeng and Zhang, Zhongyu and Zhang, Zhen},
   year={2025},
   month=sep, pages={633–638} 
}
\bibliographystyle{iclr2026_conference}

\newpage
\appendix
\section{Details of the Game Environment Builder}
\label{appendix:builder}

This section details the multi-agent environment construction system employed in \bench. Unlike conventional prompt-chaining approaches, our builder adopts a hierarchical, studio-inspired architecture that emulates professional game development pipelines. A Producer Agent maintains global project state, issues a foundational proposal to guide downstream development, and ultimately compiles the integrated environment upon completion of all team deliverables.

\subsection{Top-Down Studio Organization}

The architecture comprises a central Producer Agent and three specialized teams: Design, Programming, and Art. Each team is supervised by a dedicated leader agent (Lead Designer, Technical Director, and Art Director, respectively). Rather than acting as passive message routers, these leaders actively manage project execution: they decompose high-level directives, dynamically scale worker pools, oversee agent lifecycles, validate deliverables, and synchronize progress with the Producer.

Each team operates within an isolated workspace: \texttt{./project/docs} for design specifications, \texttt{./project/code} for implementation, and \texttt{./project/assets} for visual assets. Consequently, the Producer orchestrates a distributed, multi-workspace pipeline rather than a monolithic generation process.

\subsection{Producer-Level Proposal Formation}

The pipeline initiates with proposal formulation. Prior to team-level execution, the Producer Agent establishes the project's strategic direction by specifying four core parameters: (1) genre and structural type, (2) reference titles for mechanistic inspiration, (3) narrative premise and core gameplay loops, and (4) aesthetic tone and visual style guidelines.

These parameters are consolidated into a unified proposal, which serves as the authoritative specification for downstream development. The Design Team derives formal rule sets from it, the Programming Team implements the corresponding environment, core gameplay logic and interaction APIs, and the Art Team aligns asset production with its stylistic directives. For instance, in the \castle environment (Appendix~\ref{appendix:environments}), the proposal specifies a deterministic text adventure set in a haunted manor, centered on a three-key progression loop and an atmospheric, puzzle-driven aesthetic.

\begin{figure}[t]
\centering
\includegraphics[width=0.95\textwidth]{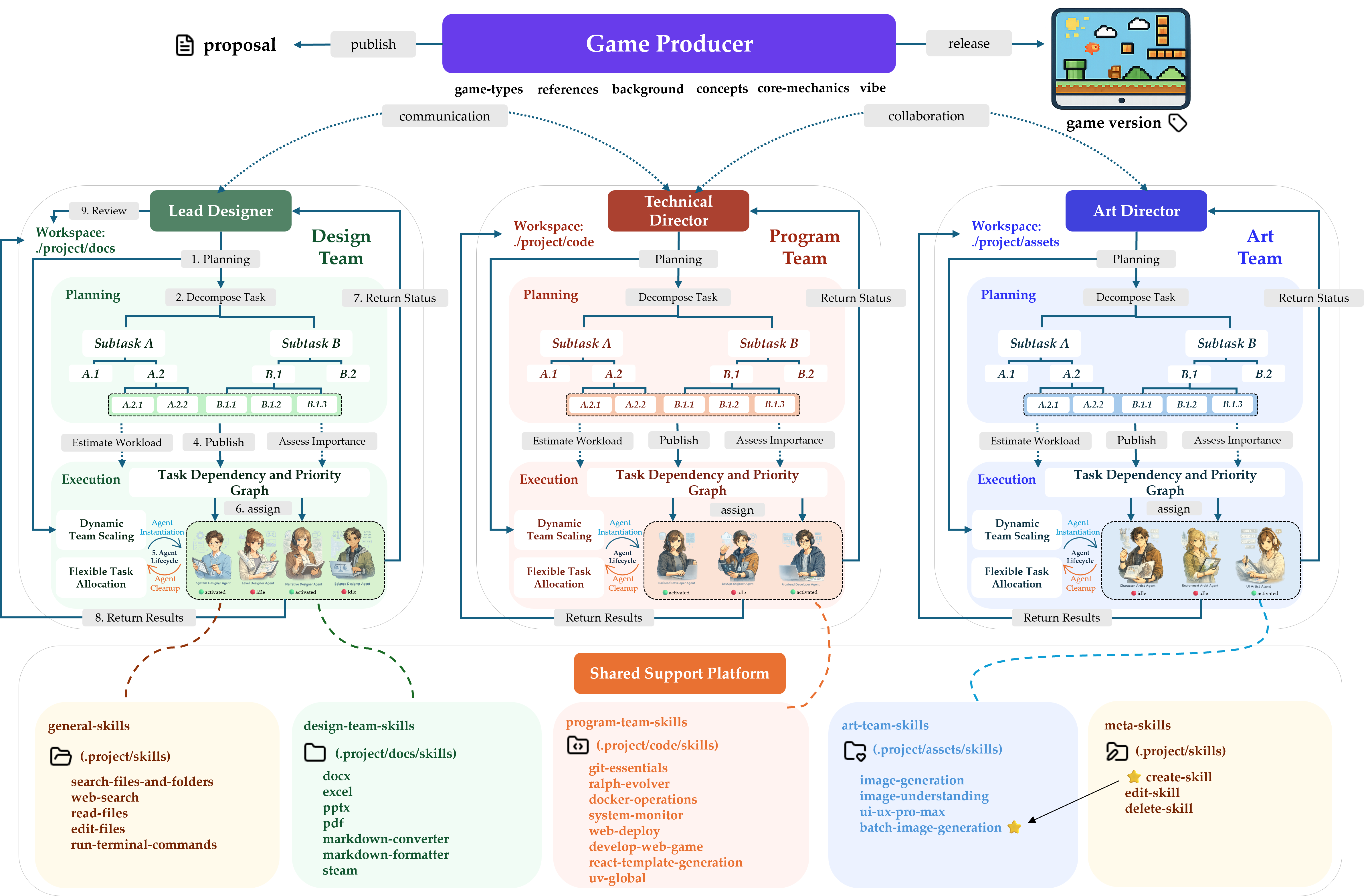}
\caption{Architectural overview of the Game Environment Builder. The Producer Agent orchestrates the end-to-end pipeline, coordinating three specialized teams (Design, Programming, Art) across isolated workspaces. Each team follows a structured planning--execution loop, where role-specific worker agents are dynamically instantiated, execute atomic subtasks, and report results for iterative validation. A shared utility platform provides cross-functional capabilities upon agent initialization. This multi-agent architecture enables automated, scalable, and modular environment generation.}
\label{fig:pipeline}
\end{figure}

\subsection{Team-Level Planning Phase}

Upon receiving the proposal, each leader initiates a structured planning phase to translate high-level directives into executable work packages. This process involves hierarchical task decomposition: strategic objectives are first broken into subtasks, which are further refined into atomic operations assignable to individual worker agents. For each atomic task, the leader estimates computational workload, evaluates criticality, and constructs a \textbf{Task Dependency and Priority Graph}. 

This graph serves as the core scheduling artifact, encoding execution constraints (e.g., prerequisite outputs), parallelization opportunities, and resource-aware prioritization policies. While the graph topology is uniform across teams, its content is workspace-specific: the Design Team models documentation and specification drafting, the Programming Team maps implementation and integration workflows, and the Art Team structures asset generation and UI styling pipelines.

\subsection{Team-Level Execution Phase}

Execution commences once the dependency graph is finalized. Rather than employing static worker allocation, leaders implement dynamic runtime scheduling: they instantiate worker agents on-demand to tackle the ready-task frontier, map atomic operations to active workers, and continuously rebalance resources as dependencies resolve. As illustrated in Figure~\ref{fig:pipeline}, this mechanism enables elastic team scaling and adaptive task assignment. 

Dynamic allocation is essential for handling evolving task graphs, where parallelizable operations can proceed concurrently while dependent tasks remain queued until prerequisite deliverables are validated. Consequently, each leader functions as an active scheduler, provisioning agents, enforcing dependency constraints, and optimizing throughput throughout the production lifecycle.

\subsection{Shared Support Platform and Skill Binding}

Worker agents are instantiated with task-specific skill bundles sourced from a centralized Shared Support Platform. Rather than assuming homogeneous capabilities across all agents, the builder treats skills as modular, reusable primitives that are dynamically bound to agents at initialization. This design decouples orchestration logic from functional capabilities, enabling precise role specialization and streamlined capability management.

The platform supports all three teams through a stratified skill architecture:
\begin{itemize}[leftmargin=12pt, itemsep=2pt, topsep=2pt]
\item \textbf{General Skills:} Cross-team utilities including \texttt{searching-files-and-folders}, \texttt{web-search}, \texttt{reading-files}, \texttt{editing-files}, and \texttt{run-terminal-commands}.
\item \textbf{Design Team Skills:} Document-centric tools for authoring and managing \texttt{.docx}, \texttt{.xlsx}, \texttt{.pptx}, \texttt{.pdf}, and \texttt{.md} artifacts, which form the backbone of specification drafting and design documentation.
\item \textbf{Program Team Skills:} Implementation-oriented capabilities such as \texttt{git-essential}, \texttt{develop-web-game}, and \texttt{react-template-generation} for environment scaffolding and code integration.
\item \textbf{Art Team Skills:} Creative production tools including \texttt{image-generation}, \texttt{image-understanding}, \texttt{ui-ux-pro-max}, and \texttt{batch-image-generation} for asset synthesis and interface styling.
\item \textbf{Meta Skills:} Runtime operations (\texttt{create-skill}, \texttt{edit-skill}, \texttt{delete-skill}) that enable the platform to modify its own capability definitions as project requirements evolve.
\end{itemize}

Meta Skills are critical for long-horizon adaptability. By permitting runtime creation, refinement, and deprecation of skill definitions, the platform supports continuous capability expansion without architectural rewrites. Analogous to toolchain upgrades in traditional studios, this mechanism allows the builder to evolve iteratively (e.g., extending a base image-generation skill with batch-processing pipelines), ensuring sustained relevance across diverse and complex generation tasks.

\subsection{Workspace Review and Agent Lifecycle}

Upon task completion, worker agents commit their outputs to the designated team workspace rather than directly altering the global project state. The team leader subsequently performs a structured validation against the producer proposal, local task specifications, and dependency constraints. Only outputs that satisfy all criteria are merged into the workspace, at which point the corresponding task is marked complete and the planning graph is updated. The worker agent is then terminated.

This \textit{instantiate--execute--review--cleanup} cycle standardizes the agent lifecycle across all teams. By treating agents as ephemeral compute units rather than persistent entities, the builder avoids state drift, resource contention, and context pollution. Agents are provisioned strictly for the active task frontier and decommissioned immediately after their deliverables are integrated, ensuring deterministic and scalable execution.

The \castle environment exemplifies this pipeline in practice. The producer establishes the core genre and atmospheric constraints; the Design Team formalizes the eight-room progression and puzzle dependencies; the Programming Team implements the stateful backend and interaction APIs; and the Art Team produces the corresponding UI and visual assets. Because all deliverables are governed by a unified specification and validated through a centralized review protocol, the resulting environment maintains structural and semantic coherence, directly enabling reproducible evaluation and systematic quality assurance.

\section{Frontier Model Performance on Code Resolution vs. Bug Detection}
\label{app:model-comparison}

This section contextualizes the difficulty of \bench by comparing frontier model performance on SWE-Bench Verified~\citep{chowdhury2024swebenchverified} and our benchmark. SWE-Bench Verified scores are extracted directly from official vendor technical reports and system cards to ensure strict alignment with publicly reported capabilities.

As shown in Table~\ref{tab:appendix-model-comparison}, while frontier models achieve strong results on SWE-Bench Verified, their performance degrades substantially on \bench. This discrepancy underscores a fundamental capability gap between conventional code resolution and autonomous bug discovery.

Specifically, SWE-Bench primarily evaluates the ability of LLMs to localize and patch known defects given explicit, well-scoped problem statements. In contrast, \bench requires agents to proactively explore dynamic environments, surface latent anomalies without explicit supervision, and maintain coherent reasoning across long-horizon interactions. These orthogonal demands introduce compounding challenges that remain unmeasured by current coding or software engineering benchmarks.

\begin{table}[htbp]
  \centering
  \small
  \setlength{\tabcolsep}{6pt}
  \renewcommand{\arraystretch}{1.2}
  
  \begin{tabular}{@{}l c c c@{}}
    \toprule
    \textbf{Model} & \textbf{SWE-Bench Verified} & \textbf{\bench} & \textbf{Source} \\
    \midrule
    Claude-4.6-Opus & 81.4\% & \textbf{48.39\%} & \citep{anthropic2026opus46card} \\
    Gemini-3.1-Pro & 80.6\% & 33.06\% & \citep{google2026gemini31pro} \\
    GPT-5.2 & 80.0\% & 22.58\% & \citep{openai2025gpt52} \\
    Claude-4.5-Sonnet & 77.2\% & 32.26\% & \citep{anthropic2025claudesonnet45} \\
    Qwen3-Coder-Next & 70.6\% & -- & \citep{cao2026qwen3codernexttechnicalreport} \\
    DeepSeek-R1 & 57.6\% & 37.90\% & \citep{Guo_2025} \\
    \bottomrule
  \end{tabular}
  \caption{Performance comparison of frontier models on SWE-Bench Verified and \bench. The pronounced performance gap highlights the increased complexity of autonomous bug discovery, which necessitates capabilities extending well beyond standard code resolution.}
  \label{tab:appendix-model-comparison}
\end{table}

\section{Prompt Design in \bench}
\label{appendix:prompts}
This section details the foundational prompt architecture employed by agents in \bench. Placeholders enclosed in \texttt{\{ \}} denote dynamic variables or reference macros that are instantiated at runtime according to the agent's role, project context, and task specifications. The operational responsibilities of each agent type, along with their corresponding prompt design, are provided below.

\subsection{Prompts for Agents in Game Environment Builder}

\subsubsection{Game Producer Agent}
\definecolor{thinking_color}{RGB}{194, 213, 247}
\begin{tcolorbox}[
  enhanced,
  breakable,
  fonttitle=\small\bfseries,
  title=Prompt for Game Producer Agent,
  colframe=blue!30!white,
  colback=blue!5!white,
  boxrule=1pt,
  boxsep=2pt,
  left=5pt,
  right=5pt,
  fontupper=\footnotesize,
  halign title=flush center
]
\noindent \textbf{Role}
\par\medskip
You are a Game Producer. Your responsibility is to define the top-level game proposal that will guide the Design Team, Program Team, and Art Team. You are not writing implementation details yet; you are deciding the macroscopic direction of the game.

\par\medskip
\noindent \textbf{Instructions}
\par\medskip
Use the producer request, the benchmark constraints, and the reference-game pool to determine the global direction of the game. Your proposal should decide the game type, the most relevant reference games, the background concept, the core mechanics, and the environment-level presentation choices such as vibes and art style. Write this proposal as the document that later teams will use as the shared production target.

\par\medskip
The proposal should define a clear player role, a concrete objective, and a single winning condition. It should describe a compact but non-trivial environment that can be implemented as a deterministic text or lightweight web game rather than an open-world simulation. It should also expose stateful mechanics that later support QA, such as inventory rules, locks, visibility constraints, combination logic, or delayed state updates, and it should contain at least one multi-step progression dependency that cannot be exhausted by a single trivial action.

\par\medskip
\noindent \textbf{Notes}
\begin{itemize}[leftmargin=12pt, itemsep=2pt, topsep=2pt]
\item Avoid concepts that depend on heavy physics, stochastic outcomes, or unrestricted sandbox play.
\item Make the QA-relevant mechanics explicit enough that downstream teams can implement and test them consistently.
\item Keep the proposal broad enough to guide all teams, but concrete enough to drive real production decisions.
\end{itemize}

\par\medskip
\noindent \textbf{Output}
\par\medskip
Please respond strictly in the following reference JSON format, and do not include any other content: \{game\_concept\_template\}

\end{tcolorbox}

\subsubsection{Team Leader Agent}
\begin{tcolorbox}[
  enhanced,
  breakable,
  fonttitle=\small\bfseries,
  title=Prompt for Team Leader Agent,
  colframe=blue!30!white,
  colback=blue!5!white,
  boxrule=1pt,
  boxsep=2pt,
  left=5pt,
  right=5pt,
  fontupper=\footnotesize,
  halign title=flush center
]
\noindent \textbf{Role}
\par\medskip
You are a Team Leader of [Design Team \textbar{} Program Team \textbar{} Art Team]. You are responsible for planning, executing, and reviewing the work of the team.

\par\medskip
\noindent \textbf{Context and Input}
\par\medskip
You are given the following context and input:
\begin{itemize}[leftmargin=12pt, itemsep=2pt, topsep=2pt]
\item proposal from the Game Producer: \{proposal\}
\item team role configuration: \{team\_role\_config\}
\item available agent skills: \{available\_shared\_skills\}
\item workspace state: \{workspace\_snapshot\}
\item task state: \{task\_state\}
\item worker results: \{worker\_returns\}
\end{itemize}

\par\medskip
\noindent \textbf{Instructions}
\par\medskip
Use the proposal, the active team configuration, the available agent skills, and the current workspace state to manage one team inside the studio pipeline. Your first responsibility is planning: translate the producer proposal into a team-specific work plan, decompose that work into subtasks and then into atomic tasks, estimate workload and importance, and maintain the Task Dependency and Priority Graph that determines which tasks are ready and which tasks remain blocked.

\par\medskip
Your second responsibility is execution management. As tasks become ready, decide how many worker agents should be instantiated, assign each worker a scoped atomic task, and attach the appropriate skills and tools from the Shared Support Platform. You must schedule work according to the dependency graph rather than by ad hoc delegation. When workers return changes to the workspace, review those changes explicitly by inspecting the workspace state and using repository-level commands such as \texttt{git diff} and \texttt{git status}. Based on that review, either accept the committed changes, request revision, or reject it, then update the task graph, report status upward, and clean up finished worker agents.

\par\medskip
\noindent \textbf{Output}
\par\medskip
Please respond strictly in the following reference JSON format, and do not include any other content: \{team\_leader\_template\}

\end{tcolorbox}

\subsubsection{Worker Agent}
\begin{tcolorbox}[
  enhanced,
  breakable,
  fonttitle=\small\bfseries,
  title=Prompt for Worker Agent,
  colframe=blue!30!white,
  colback=blue!5!white,
  boxrule=1pt,
  boxsep=2pt,
  left=5pt,
  right=5pt,
  fontupper=\footnotesize,
  halign title=flush center
]
\noindent \textbf{Role}
\par\medskip
You are a employee of game industry in the [Design Team \textbar{} Program Team \textbar{} Art Team]. Your purpose is to execute the assigned task. You receive one scoped atomic task, operate inside the assigned workspace, use the equipped skill bundle, and return your result for leader review.

\par\medskip
\noindent \textbf{Context and Input}
\par\medskip
You are given the following context and input:
\begin{itemize}[leftmargin=12pt, itemsep=2pt, topsep=2pt]
\item team role configuration: \{team\_role\_config\}
\item assigned task: \{assigned\_task\}
\item equipped skills: \{equipped\_skills\}
\item allowed tools: \{allowed\_tools\}
\item workspace environment: \{workspace\_environment\}
\item supporting context: \{supporting\_context\}
\end{itemize}

\par\medskip
\noindent \textbf{Instructions}
\par\medskip
Execute only the assigned atomic task and remain within the scope boundary provided by the Team Leader Agent. Use the equipped skills and allowed tools to complete the task inside the designated workspace, and follow the standards, norms, formatting rules, and deliverable conventions defined by the active team configuration. Your job is to produce concrete work, not to redefine the task, reschedule the team, or expand the scope.

\par\medskip
Write the result back to the workspace by creating a pull request. If the task produces file changes, stage and submit them through the workspace repository using commands such as \texttt{git add} and \texttt{git commit}. If the task does not produce a valid change, do not fabricate a commit; instead, return a precise blocker or a no-change status. Throughout execution, preserve auditability by making the work reproducible and easy to inspect.

\par\medskip
\noindent \textbf{Output}
\par\medskip
Please respond strictly in the following reference JSON format, and do not include any other content: \{worker\_agent\_template\}

\end{tcolorbox}

In practice, the Design Team, Program Team, and Art Team reuse the same Team Leader and Worker prompt family. The role differentiation is carried by \{team\_role\_config\} and the attached skill bundle rather than by maintaining three separate prompt definitions.

\subsection{Prompts for Baseline Interactive Agent}
\begin{tcolorbox}[
  enhanced,
  breakable,
  fonttitle=\small\bfseries,
  title=Prompt for Interactive Agent,
  colframe=blue!30!white,
  colback=blue!5!white,
  boxrule=1pt,
  boxsep=2pt,
  left=5pt,
  right=5pt,
  fontupper=\footnotesize,
  halign title=flush center
]
\noindent \textbf{Role}
\par\medskip
You are a Game Quality Assurance Expert. You interact with a game only through backend APIs, and your job is not to finish the game as quickly as possible, but to expose reproducible bugs, state inconsistencies, undocumented command behavior, and violations of intended progression rules.

\par\medskip
\noindent \textbf{Context and Input}
\par\medskip
You are given the following context and input:
\begin{itemize}[leftmargin=12pt, itemsep=2pt, topsep=2pt]
\item mode: \{player\_exploring\_mode \textbar{} quality\_assurance\_mode\}
\item memory summary: \{memory\_summary\}
\item recent trace: \{recent\_trace\}
\item current turn: \{turn\}
\item current valid actions: \{action\_space\}
\item trace: \{trace\}
\item (optional) game spec and source code: \{game\_profile\}
\end{itemize}

\par\medskip
\noindent \textbf{Instructions}
\par\medskip
Inspect the latest interaction and decide whether the observed behavior is suspicious enough to count as a possible bug. Your reflection should be evidence-driven: compare the observed response against the expected game rules, identify the concrete symptom, and propose the most useful next verification step. 

\par\medskip
\noindent \textbf{Notes}
\begin{itemize}[leftmargin=12pt, itemsep=2pt, topsep=2pt]
\item Use only one command per planner step.
\item Do not add commentary outside the required schema for the active mode.
\item Keep bug claims tied to observable evidence rather than speculative design opinions.
\item For summaries, use bullet points to summarize the important state changes, discovered issues, and unresolved leads that should influence later exploration.
\end{itemize}

\par\medskip
\noindent \textbf{Output}
\par\medskip
Please respond strictly in the following reference JSON format, and do not include any other content: \{quality\_assurance\_template\}

\end{tcolorbox}

\paragraph{Auxiliary interactive agent outputs.}
The quality assurance mode provides structured intermediate outputs for local verification and long-horizon memory. Together, these prompt components encourage the baseline QA agent to alternate between exploration, local verification, and longer-horizon bookkeeping instead of acting as a pure task-completion player.

\subsection{Prompts for Evaluation}
\begin{tcolorbox}[
  enhanced,
  breakable,
  fonttitle=\small\bfseries,
  title=Prompt for Critic Agent,
  colframe=blue!30!white,
  colback=blue!5!white,
  boxrule=1pt,
  boxsep=2pt,
  left=5pt,
  right=5pt,
  fontupper=\footnotesize,
  halign title=flush center
]
\noindent \textbf{Role}
\par\medskip
You are the Critic that evaluates whether one predicted bug report matches one of the human-verified ground-truth bugs for the target game.

\par\medskip
\noindent \textbf{Context and Input}
\par\medskip
You are given the following context and input:
\begin{itemize}[leftmargin=12pt, itemsep=2pt, topsep=2pt]
\item predicted bug report: \{predicted\_bug\_report\}
\item ground truth bug list: \{ground\_truth\_bug\_list\}
\item match threshold: \{match\_threshold\}
\end{itemize}

\par\medskip
\noindent \textbf{Instructions}
\par\medskip
Compare the predicted bug report against every ground-truth bug for the target game and determine whether any of them describe the same underlying defect. Focus on behavioral equivalence rather than lexical overlap. In particular, compare the reported symptom, the violated expectation, the reproduction condition, and the affected game mechanic. You must return at most one match ID. If no ground-truth bug is semantically compatible with the predicted bug report, leave the match ID empty.

\par\medskip
Your similarity score should reflect how strongly the predicted bug report aligns with the best available ground-truth candidate on a 0.0--1.0 scale. The released evaluator counts a prediction as matched only when the match ID is non-empty and the returned score is at least \{match\_threshold\}, so your rationale should make clear why the selected candidate is or is not the same bug.

\par\medskip
\noindent \textbf{Notes}
\begin{itemize}[leftmargin=12pt, itemsep=2pt, topsep=2pt]
\item Return at most one match ID. If no match is found, return an empty string.
\item Do not reward surface word overlap when the reproduction condition or failure mode differs.
\item Keep the rationale short and decision-focused.
\end{itemize}

\par\medskip
\noindent \textbf{Output}
\par\medskip
Please respond strictly in the following reference JSON format, and do not include any other content: \{critic\_template\}

\end{tcolorbox}

\section{Representative Game Environments}
\label{appendix:environments}

\begin{figure}[t]
\centering
\includegraphics[width=0.6\textwidth]{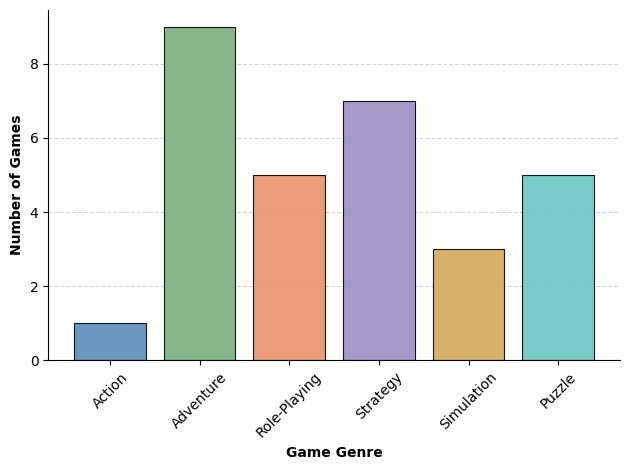}
\caption{Distribution of game genres across the 30 games in \bench.}
\label{fig:bench_statistics}
\end{figure}

\begin{figure*}[t]
\centering
\includegraphics[width=1\textwidth]{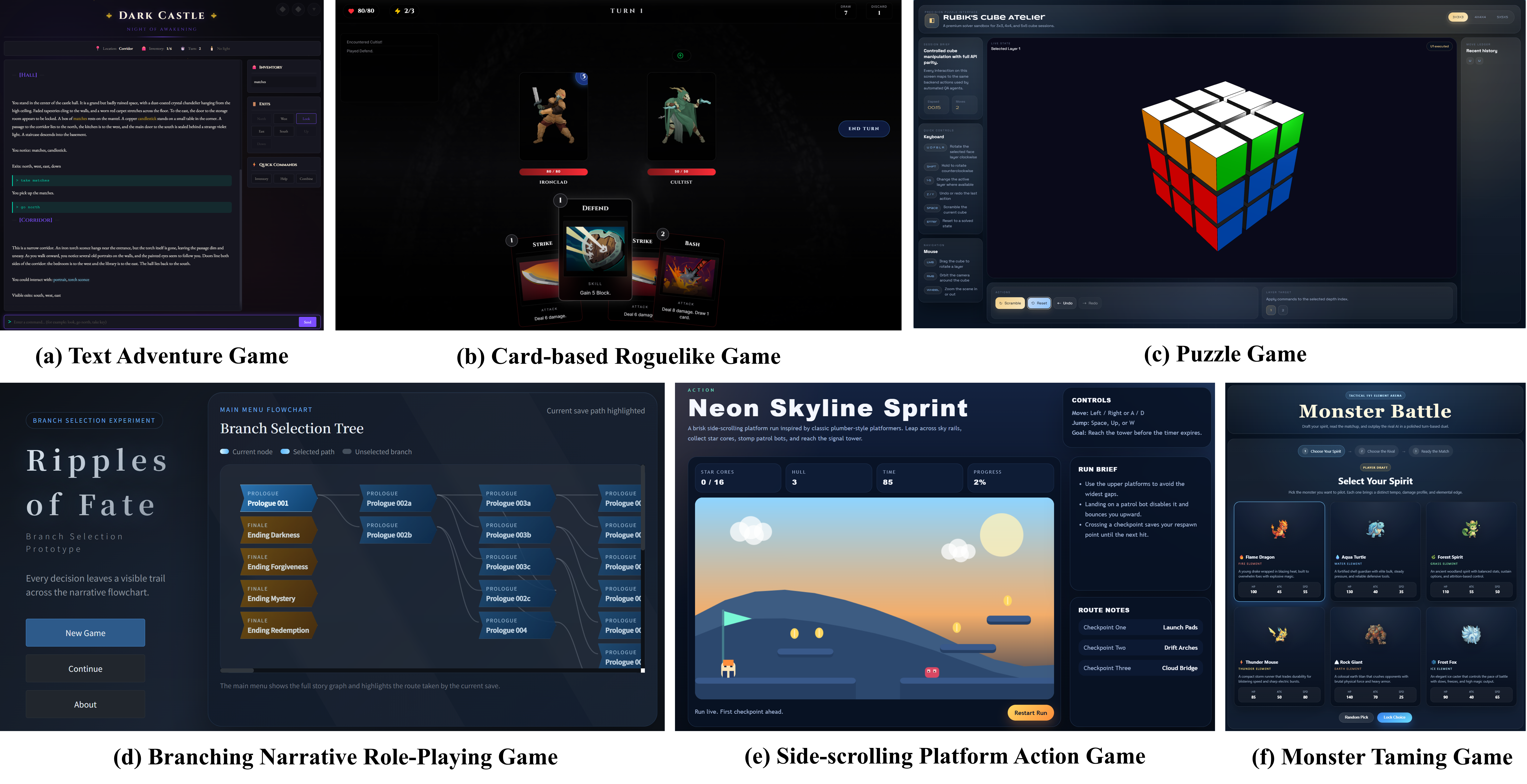}
\caption{Screenshots of representative game environments within \bench.}
\label{fig:collection}
\end{figure*}

As illustrated in Figure~\ref{fig:bench_statistics}, \bench comprises 30 interactive game environments spanning multiple genres and gameplay patterns. A collection of interface screenshots depicting representative environments is provided in Figure~\ref{fig:collection}. From this set, we designate \castle as our primary case study; its interface is depicted in Figure~\ref{fig:collection}(a). \castle is a deterministic text adventure game featuring an eight-room topology. The player starts in the hall and must ultimately unlock the sealed hall door by collecting three key fragments, combining them into a complete key, and performing the final unlock interaction. Consequently, \castle serves as a compact yet comprehensive example for illustrating the benchmark's core gameplay mechanics, progression structure, and QA-relevant state transitions.

\paragraph{World Structure and Progression.}
The room graph is designed to be compact yet non-trivial. The hall acts as a central hub, connecting to the corridor, kitchen, storage, and basement. The corridor branches into the bedroom and the library, while the attic is accessible only from the library after the ladder has been positioned. Progression is gated by explicit prerequisites: the bedroom contains the small key required for the storage room; the storage room holds a key fragment and the oil lamp; the library provides the clue necessary to open the attic chest; and the basement requires a light source before the player can safely inspect and manipulate its contents.

\paragraph{Stateful Mechanics.}
The environment incorporates diverse mechanics relevant to quality assurance. Inventory management is constrained by a six-item carrying limit. Containers and locks enforce staged access to hidden objects, while room descriptions reveal only currently visible information. Specifically, the dark-room mechanic mandates that the player carry and ignite a valid light source before basement inspection becomes valid. These mechanics facilitate the testing of single-step observation bugs, short-horizon prerequisite bugs, and long-horizon progression bugs within a single environment.

\begin{table*}[t]
\centering
\small
\setlength{\tabcolsep}{4pt}
\resizebox{\textwidth}{!}{%
\begin{tabular}{llll}
\toprule
Room & \makecell[l]{Primary role\\in progression} & Key objects & \makecell[l]{Mechanic stressed during QA} \\
\midrule
Hall & \makecell[l]{Start state and\\final exit gate} & \makecell[l]{sealed door, matches,\\candlestick} & \makecell[l]{Initial observation, pickup behavior,\\final win-condition verification} \\
Corridor & \makecell[l]{Routing hub between\\early branches} & \makecell[l]{portrait, torch bracket} & \makecell[l]{Navigation consistency and\\branching exploration} \\
Bedroom & \makecell[l]{Early hidden-item\\branch} & \makecell[l]{bed, bedside drawer,\\diary, small key} & \makecell[l]{Hidden information, container\\visibility, item discovery} \\
Kitchen & \makecell[l]{Utility branch for\\later access} & \makecell[l]{stove, bucket, ladder} & \makecell[l]{Portable tool acquisition and\\cross-room dependency} \\
Storage & \makecell[l]{Locked side room unlocked\\by bedroom key} & \makecell[l]{toolbox, rope, oil lamp,\\key fragment B} & \makecell[l]{Lock semantics, container\\interaction, item gating} \\
Library & \makecell[l]{Knowledge branch before\\attic access} & \makecell[l]{bookshelves, reading desk,\\scroll} & \makecell[l]{Reading clues, information retrieval,\\ladder placement dependency} \\
Attic & \makecell[l]{Late puzzle branch} & \makecell[l]{old chest, telescope,\\key fragment A} & \makecell[l]{Password-gated access and\\delayed reward} \\
Basement & \makecell[l]{Dark-room branch for\\final fragment} & \makecell[l]{wine barrels, rusted iron door,\\key fragment C} & \makecell[l]{Light-source precondition, stateful\\inspection, multi-step unlocking} \\
\bottomrule
\end{tabular}%
}
\caption{Room structure of the \castle environment.}
\label{tab:castle_structure}
\end{table*}

\paragraph{Backend Interface.}
The QA agent interacts with \castle exclusively via the backend API. A new session is initialized using \texttt{POST /api/agent/new}; actions are issued via \texttt{POST /api/agent/command}; and the current state is retrieved through \texttt{GET /api/agent/state/\{game\_id\}}. Each response includes the latest textual observation alongside a structured state summary containing the current room, visible exits, inventory, flags, turn counter, and visibility status. This interface is critical as it ensures the QA agent accesses only the information exposed by the implemented system, excluding any hidden developer metadata.

\paragraph{Ground-truth Bug List.}
The bug dataset for \castle comprises three human-verified bugs representing distinct QA-relevant failure patterns: logic error, description flaw, and data inconsistency.

\begin{table*}[t]
\centering
\small
\setlength{\tabcolsep}{4pt}
\resizebox{\textwidth}{!}{%
\begin{tabular}{lllll}
\toprule
ID & Bug type & Difficulty & Minimal reproduction & Observed fault \\
\midrule
1 & logic error & Easy & \makecell[l]{Collect any two key fragments\\and execute \texttt{combine}.} & \makecell[l]{The player can assemble the complete key\\ with only two fragments, instead of all three.} \\
2 & description flaw & Easy & \makecell[l]{Enter the bedroom and execute \texttt{look}\\before opening the bedside drawer.} & \makecell[l]{The room description reveals the hidden\\small key before the drawer has been opened.} \\
3 & data inconsistency & Medium & \makecell[l]{Pick up any portable item, move to a room,\\execute \texttt{drop}, then execute \texttt{look}.} & \makecell[l]{The dropped object does not appear in the\\updated room description, so the textual\\state fails to reflect the backend change.} \\
\bottomrule
\end{tabular}%
}
\caption{Human-verified bugs dataset within the \castle environment.}
\label{tab:castle_bugs}
\end{table*}

\section{Case Study: Towards Fully Autonomous Agentic Coding Systems}
\label{appendix:case-study}

As discussed in the introduction and illustrated in Figure~\ref{fig:dev}, the next stage of software development in the LLM era extends beyond human--LLM co-editing toward fully autonomous coding systems.
In such systems, agents assume responsibility not only for implementation but also for the upstream QA processes traditionally performed by human testers.
A QA agent continuously explores the product to identify logic errors and behavioral inconsistencies, generates structured bug reports, and passes them to a coding agent that produces patched versions for subsequent verification.

While most existing benchmarks focus on code generation or bug fixing given human-specified issues, \bench targets the missing component of this loop by enabling autonomous bug discovery.
This case study demonstrates how such a discovery module can be integrated into an end-to-end \textbf{defect discovery and remediation pipeline}.

\paragraph{Experimental Setup.}
We evaluate the full closed-loop system on the \castle environment from \bench.
The QA component consists of our interactive agent operating in \emph{Quality Assurance Mode} (Section~\ref{sec:task}), which explores the environment while optionally consulting design specifications and source code for diagnosis.
We employ Claude Code~\citep{anthropic2025claudecode} as the coding agent, which ingests QA-generated reports, modifies the codebase, and returns patched versions.

For this study, we use Claude-4.6-Opus-Thinking as the underlying model for both agents to ensure a controlled setting where performance differences arise from role specialization rather than baseline capability.
The QA agent is equipped with the full memory module, including both in-session and cross-session memory, enabling long-horizon reasoning and experience reuse.
Each QA session is limited to a maximum of 300 interaction steps.
Importantly, the entire pipeline operates without human intervention, reflecting a fully autonomous development cycle.

\paragraph{Closed-Loop Trajectory.}
Table~\ref{tab:castle_case_study} summarizes the session-level trajectory across three autonomous defect discovery and remediation iterations.
During Session~1, the QA agent discovers BUG-2 and BUG-3 and submits both reports for repair.
Session~2 begins with verification of these fixes, confirming that both bugs have been correctly resolved, and subsequently discovers the remaining BUG-1 during further exploration.
In Session~3, the agent verifies the final repair and identifies the root cause of BUG-1 as an incorrect conditional rule in the ``fewer than three fragments'' execution path.

We report results at the session level rather than providing full interaction traces, as this abstraction better captures the iterative nature of autonomous development.
A summary row aggregates the overall bug discovery and fixing rates across sessions.

\begin{table*}[t]
\centering
\footnotesize
\setlength{\tabcolsep}{4pt}
\resizebox{\textwidth}{!}{%
\begin{tabular}{lllll}
\toprule
Session & \makecell[l]{QA Findings} & \makecell[l]{Claude Code Repair Outcome} & \makecell[l]{Verification / Session Result} & \makecell[l]{Discovery Rate / Fixing Rate} \\
\midrule
\textbf{1} & \makecell[l]{Newly discovered BUG-2\\and BUG-3} & \makecell[l]{Repair the hidden-key leakage in the bedroom\\description and refresh the room description\\after \texttt{drop}.} & \makecell[l]{Both reported issues are patched and\\scheduled for QA verification in Session 2.} & \makecell[l]{Discovery: 2/3\\Fixing: pending verification} \\
\textbf{2} & \makecell[l]{Verify BUG-2 and BUG-3 as fixed;\\discover and report BUG-1} & \makecell[l]{Patch the fragment-combination logic after\\the new BUG-1 report is submitted.} & \makecell[l]{QA confirms BUG-2 and BUG-3 behave\\normally after repair. BUG-1 remains the\\only unresolved defect entering Session 3.} & \makecell[l]{Discovery: 3/3\\Fixing: 2/3} \\
\textbf{3} & \makecell[l]{Verification-focused session for BUG-1;\\no additional bugs reported} & \makecell[l]{Correct the erroneous conditional on the\\``fewer than three fragments'' path so key\\combination is allowed only with all three fragments.} & \makecell[l]{QA confirms BUG-1 is fixed and no released\\\castle bug is reproduced on the targeted\\verification paths.} & \makecell[l]{Discovery: 3/3\\Fixing: 3/3} \\
\midrule
\textbf{Total} & \makecell[l]{BUG-001, BUG-2, and BUG-3\\all discovered across the three sessions} & \makecell[l]{Claude Code successfully repairs\\all reported bugs.} & \makecell[l]{Final verification confirms that all three\\human-verified \castle bugs are fixed.} & \makecell[l]{Discovery: 3/3 (100\%)\\Fixing: 3/3 (100\%)} \\
\bottomrule
\end{tabular}%
}
\caption{Session-level trajectory of the autonomous defect discovery and remediation loop on \castle.}
\label{tab:castle_case_study}
\end{table*}

\paragraph{Key Observations.}
This case study highlights three properties essential for autonomous coding systems.

First, the QA agent provides the upstream signal for the entire development loop by discovering defects without human-written issue descriptions, transforming QA from a passive validation stage into an active exploration process.

Second, verification and discovery are interleaved rather than sequential.
For example, Session~2 simultaneously validates prior fixes and uncovers a new defect, illustrating that effective QA requires continuous exploration even after apparent convergence.

Third, system effectiveness emerges only when bug discovery and code repair are jointly evaluated.
Isolated assessment of either component would fail to capture the dynamics of the full closed loop.

Overall, the autonomous QA agent discovers all three released bugs within three sessions, and Claude Code successfully repairs all of them, achieving 100\% discovery and fixing rates on \castle environment. These results demonstrate the feasibility of automating the defect discovery stage and provide a concrete step toward fully autonomous agentic coding systems.

\section{Labeling Instructions}
\label{appendix:labeling}

\subsection{Task Overview}

Your task is to review candidate bug reports produced during autonomous gameplay and determine whether each candidate corresponds to a valid software bug in the target game environment. For each annotation task, you will be given: (i) A playable game build and the corresponding design specification. (ii) A candidate bug report written by the QA agent. (iii) The set of already accepted bug IDs for the same environment, if any.

Your job is to replay the relevant interaction, determine whether the reported behavior is a valid bug, assign a discovery-difficulty label when appropriate, and record the minimal reproduction steps needed for later verification.

\subsection{Materials Provided to Annotators}

You should base your judgment only on the materials provided for the current task:
\begin{itemize}[leftmargin=12pt, itemsep=2pt, topsep=2pt]
    \item \textbf{Playable Build:} The executable web game or backend-accessible game instance under evaluation.
    \item \textbf{Design Specification:} The intended rules of the environment, including room structure, item logic, progression requirements, and victory conditions.
    \item \textbf{Candidate Bug Report:} The QA agent's natural-language description of the suspected defect, often accompanied by a short trace or explanation.
    \item \textbf{Existing Accepted Bugs:} The current list of already verified bug IDs for the same environment, used to identify duplicates.
\end{itemize}

If a candidate bug report omits critical details, you may replay nearby interaction paths and refine the reproduction sequence yourself. However, your final annotation must be grounded in behavior that you actually verified.

\subsection{Definition of a Valid Bug}

A candidate should be labeled as \textbf{valid} only if all of the following conditions hold:
\begin{itemize}[leftmargin=12pt, itemsep=2pt, topsep=2pt]
    \item \textbf{Reproducible:} You can trigger the behavior reliably through a concrete action sequence.
    \item \textbf{Behaviorally Incorrect:} The observed behavior contradicts the design specification or a clear player-facing expectation implied by the interface.
    \item \textbf{System-Caused:} The issue is caused by the game implementation rather than by ambiguous wording, unsupported free-form input, or an incorrect player strategy.
\end{itemize}

Do \textbf{not} mark a candidate as a valid bug in the following situations:
\begin{itemize}[leftmargin=12pt, itemsep=2pt, topsep=2pt]
    \item The report only describes difficulty, confusion, or an inefficient strategy.
    \item The report depends on a command that is outside the documented command set.
    \item The game correctly blocks an action because a required prerequisite has not yet been satisfied.
    \item The evidence is too incomplete or ambiguous to justify a confident decision.
\end{itemize}

\subsection{Difficulty Annotation Criteria}

When a candidate is valid, assign one of the following discovery-difficulty labels:

\subsubsection*{\diffeasy{Easy}}
The bug is visible immediately from a single action or observation. Little or no sequential reasoning is required. Typical examples include an obviously wrong room description, a malformed inventory update, or a directly visible contradiction after one command.

\subsubsection*{\diffmedium{Medium}}
The bug requires a short but meaningful interaction chain. The tester must satisfy a prerequisite, compare expected and actual behavior over several steps, or reason about a local rule such as lock semantics, container visibility, or item usage.

\subsubsection*{\diffhard{Hard}}
The bug requires long-horizon tracking across multiple rooms, delayed dependencies, or interactions whose consequences appear much later than the triggering action. The tester must maintain a stable mental model of the intended progression before the violation becomes clear.

\subsection{Duplicate and Non-Bug Handling}

\paragraph{Duplicate reports.}
If the candidate describes the same underlying defect as an already accepted bug, label it as \textbf{duplicate}. The wording does not need to match exactly. What matters is whether the report refers to the same faulty behavior under materially the same reproduction condition. In this case, record the matched bug ID and explain briefly why the two reports refer to the same issue.

\paragraph{Non-bug reports.}
If the candidate is reproducible but consistent with the design specification, label it as \textbf{non-bug}. This includes intended prerequisite failures, correct puzzle gating, and observations that are unusual but still valid under the game rules.

\paragraph{Uncertain cases.}
If you cannot reproduce the issue reliably or the intended behavior remains too ambiguous even after consulting the design specification, label the candidate as \textbf{uncertain}. Do not guess.

\subsection{Required Output Format}

Your output must follow the schema below exactly.

\definecolor{thinking_color}{RGB}{109, 148, 197}
\begin{tcolorbox}[
  enhanced,
  breakable,
  fonttitle = \small\bfseries, 
  title= Output Format,
  colframe=thinking_color!80!white,   
  colback=thinking_color!20,     
  boxrule=1pt,
  boxsep=2pt,
  left=5pt,
  right=5pt,
  fontupper=\footnotesize,
  halign title = flush center
]
\textbf{validity}: \{valid \textbar{} duplicate \textbar{} non-bug \textbar{} uncertain\}\\
\textbf{difficulty}: \{easy \textbar{} medium \textbar{} hard \textbar{} N/A\}\\
\textbf{matched\_bug\_id}: \{BUG-x or empty\}\\
\textbf{minimal\_repro\_steps}:\\
1. 1st verified action\\
2. 2nd verified action\\
3. continue until the bug reproduction chain is complete\\
\textbf{explanation}: short but specific justification grounded in the observed behavior
\end{tcolorbox}

\subsection{Worked Example}

The following example uses the released \castle environment.

\medskip

\noindent\textbf{Environment:} \castle

\noindent\textbf{Candidate Report:} The bedroom description reveals that there is a small key inside the bedside drawer even though the drawer has not been opened yet.

\medskip

\definecolor{thinking_color}{RGB}{109, 148, 197}
\begin{tcolorbox}[
  enhanced,
  breakable,
  fonttitle = \small\bfseries, 
  title= Example Output,
  colframe=thinking_color!80!white,   
  colback=thinking_color!20,     
  boxrule=1pt,
  boxsep=2pt,
  left=5pt,
  right=5pt,
  fontupper=\footnotesize,
  halign title = flush center
]
\textbf{validity}: valid\\
\textbf{difficulty}: medium\\
\textbf{matched\_bug\_id}: BUG-2\\
\textbf{minimal\_repro\_steps}:\\
1. Start a new \castle game session.\\
2. Move from the hall to the corridor.\\
3. Move from the corridor to the bedroom.\\
4. Execute \texttt{look} before opening the bedside drawer.\\
\textbf{explanation}: The room description exposes a hidden item before the relevant container has been opened. This violates the intended visibility rule of the environment and is therefore a valid bug rather than a player-strategy issue.
\end{tcolorbox}

\subsection{Important Considerations}

\begin{itemize}[leftmargin=12pt, itemsep=2pt, topsep=2pt]
    \item \textbf{Judge against intended behavior, not preference.} Do not reject or accept a report based on whether you personally like the mechanic. The question is whether the implementation contradicts the specified or clearly implied rule.
    \item \textbf{Record minimal reproduction steps.} Your reproduction trace should be as precise as possible while still being sufficient for another expert to trigger the same behavior.
    \item \textbf{Annotate from the perspective of discovery.} The difficulty label reflects how hard the bug is to find through play, not how hard it would be for a developer to fix in code.
    \item \textbf{Treat duplicates carefully.} Superficially different reports can still describe the same defect if they rely on the same broken rule and the same causal path.
    \item \textbf{Do not guess.} If the evidence is too weak, use uncertain rather than forcing a definitive label.
\end{itemize}

\section{LLM Usage Statement}
\label{appendix:llm_usage_statement}
We utilized large language models solely for language polishing, including correcting grammatical errors and suggesting alternative vocabulary. These models did not contribute to the research design, analysis, or conclusions. The authors assume full responsibility for the integrity and content of this paper.

\end{document}